\documentclass{aa}  
\usepackage{graphicx}
\usepackage{txfonts}
\usepackage[colorlinks=true,allcolors=blue]{hyperref}
\usepackage{subcaption}
\usepackage{adjustbox}
\usepackage{verbatim}
\usepackage{multirow}
\begin{document}

\title{AGN X-ray coronae at high luminosity: A broadband study of RBS 229 and PG~1407+265}

\author{
Alessandro Leonardo Lai\inst{1} \and 
Stefano Bianchi\inst{1} \and 
Andrea Marinucci\inst{2} \and 
Enrico Piconcelli\inst{3} \and 
Francesco Ursini\inst{1} \and 
Elena Bertola\inst{6} \and
Giorgio Lanzuisi\inst{7} \and
Pierre-Olivier Petrucci\inst{4} \and
Daniel Stern\inst{5} \and
Cristian Vignali\inst{8} \and
Luca Zappacosta\inst{3}
}

\institute{
Dipartimento di Matematica e Fisica, Università degli Studi Roma Tre, via della Vasca Navale 84, I-00146 Roma, Italy\\
\email{alessandroleonardo.lai@uniroma3.it}
\and
ASI -- Agenzia Spaziale Italiana, Via del Politecnico snc, 00133 Roma, Italy
\and
INAF -- Osservatorio Astronomico di Roma, Via Frascati 33, 00078 Monte Porzio Catone, Italy
\and
Univ. Grenoble Alpes, CNRS, IPAG, 38000 Grenoble, France
\and
Jet Propulsion Laboratory, California Institute of Technology, Pasadena, CA 91109, USA
\and
INAF – Osservatorio Astrofisico di Arcetri, largo E. Fermi 5, 50127 Firenze, Italy
\and
INAF – Osservatorio di Astrofisica e Scienza dello Spazio di Bologna, Via Gobetti, 93/3, I-40129 Bologna, Italy
\and
Dipartimento di Fisica e Astronomia, Alma Mater Studiorum, Università degli Studi di Bologna, Via Gobetti 93/2, 40129 Bologna, Italy
}

\abstract{The X-ray emission of active galactic nuclei (AGNs) is commonly interpreted as arising from Comptonisation in a hot corona located above the accretion disc. In many sources, a prominent soft X-ray excess is also observed, often explained within the two-corona scenario, where a warm Comptonising layer produces the soft excess, while a hotter corona generates the hard X-ray continuum. However, the physical properties of these coronal components remain poorly constrained in very luminous quasars.}
{We investigate the structure and physical properties of the X-ray-emitting coronae in two luminous radio-quiet quasars, PG~1407+265 ($z=0.94$) and RBS~229 ($z=0.334$), and test whether the two-corona scenario provides a consistent description of their broadband spectra.}
{We performed a broadband spectral analysis combining \textit{XMM-Newton} EPIC and Optical Monitor data with simultaneous \textit{NuSTAR} observations and long-term \textit{Swift} monitoring over months to years. The spectra were modelled using Comptonisation models to characterise the properties of both the warm and hot coronae.}
{Both quasars are X-ray-loud with respect to the standard $\alpha_{\rm OX}$--$L_{\rm UV}$ relation, with PG~1407+265 exhibiting the strongest deviation. Their broadband spectra are well reproduced by a two-corona configuration, consisting of a warm, optically thick Comptonising component that accounts for the soft excess and a hot corona responsible for the hard X-ray continuum. Both sources show high energy cut-offs in the range $\sim70-80$ keV. RBS~229 exhibits moderate reflection and a weak neutral Fe K$\alpha$ line, whereas no significant reflection features were detected in PG~1407+265. The soft excess is present in both quasars, but displays a markedly different behaviour for each: the warm-corona parameters in RBS~229 remain consistent across epochs, while PG~1407+265 displays strong variability, producing an extreme soft-excess episode extending up to $\sim3$ keV in the rest frame.}
{These results provide new constraints on the properties of AGN coronae in the high-luminosity regime, which remains relatively poorly explored in observations. 
Both quasars show high energy cut-offs that are broadly consistent with the trends observed in terms of luminosity and accretion rates among AGN samples and safely below the runaway pair-production limit. 
No clear dependence of the warm-corona properties on the luminosity or the accretion rate was found, as both quasars show similar $kT$, $\tau$, and soft-excess strengths despite their different regimes, with variations seen mainly in terms of the spectral slope and variability. The strong variability of the warm component in PG~1407+265, compared to the stability of RBS~229, supports a scenario in which the warm corona responds to accretion-flow changes, while remaining within a narrow range of physical conditions.}


\maketitle
\nolinenumbers

\section{Introduction}
\begin{table*}[ht!]
\caption{Summary of the \textit{XMM-Newton} and \textit{NuSTAR}
observations.}
\label{tab:obs_all}
\centering

\begin{tabular}{llccccl}
\hline
Name & ObsID & Date & Exp. (ks) & Ext. radius (arcsec) &
TET (ks) & Filters \\
\hline
\multicolumn{7}{c}{\textbf{PG~1407+265}} \\
PG\_NuSTAR\_2025B
    & 61001016004 & 22-01-2025 & 142
    & 49 & 285 & -- \\
PG\_NuSTAR\_2025A
    & 61001016002 & 17-01-2025 & 88
    & 54 & 175 & -- \\
PG\_XMM\_2025
    & 0935790201 & 18-01-2025 & 15
    & pn=35; MOS$_{1,2}$=32,31 & 35 & UVM2, UVW1 \\
PG\_XMM\_2001B
    & 0092850501 & 22-12-2001 & 39
    & pn=36; MOS$_{1,2}$=40 & 42 & UVW2 \\
PG\_XMM\_2001A
    & 0092850101 & 23-01-2001 & 61
    & pn=27; MOS$_{1,2}$=40 & 69 & UVW2 \\
\hline
\multicolumn{7}{c}{\textbf{RBS~229}} \\
RBS\_NuSTAR\_2025B
    & 61001015006 & 16-01-2025 & 68
    & 54 & 130 & U (\textit{Swift}) \\
RBS\_NuSTAR\_2025A
    & 61001015004 & 05-01-2025 & 62
    & 54 & 118 & U (\textit{Swift}) \\
RBS\_XMM\_2024
    & 0935790101 & 31-12-2024 & 20
    & pn=20; MOS$_{1,2}$=40,39 & 23 & UVM2, UVW1 \\
RBS\_NuSTAR\_2024
    & 61001015002 & 30-12-2024 & 150
    & 54 & 296 & -- \\
RBS\_XMM\_2015
    & 0744450301 & 29-01-2015 & 137
    & pn=29; MOS$_{1,2}$=34,32 & 141
    & UVW2, UVM2, UVW1, U \\
\hline
\end{tabular}

\tablefoot{
The columns list the observation name, observation identifier, start
date, total exposure time, source-extraction radius, total elapsed time
(TET), and available optical monitor filters. For \textit{NuSTAR}
observations that were not simultaneous with the corresponding
\textit{XMM-Newton} observations. The associated \textit{Swift} data
are also listed.
}
\end{table*}
Active galactic nuclei (AGNs) represent one of the most efficient engines of energy production in the Universe, powered by accretion onto supermassive black holes (SMBHs). Their X-ray emission is generally interpreted as arising from a hot, optically thin corona that Comptonises seed photons from the accretion disc, producing a power-law continuum with photon index of $\Gamma \simeq 1.7$--$2.3$ \citep{HaardtMaraschi1991,Fabian2015}.
In the framework of the two-phase disc-corona model, the coronal temperature is regulated by radiative equilibrium between the hot plasma and the underlying accretion disc \citep{haardt1993x}. Hard X-ray photons emitted by the corona are intercepted and reprocessed by the disc into soft radiation, which, in turn, cools the corona via inverse Compton scattering. This radiative coupling establishes a self-consistent equilibrium in which the electron temperature adjusts to maintain a balance between the soft and hard luminosities of the system.

In pair-regulated models, electron--positron pair production prevents the coronal temperature from exceeding a critical limit, effectively acting as a ``thermostat'' through pair creation and annihilation processes \citep{Fabian2015}.
Measurements of the photon index ($\Gamma$) and high energy cut-off ($E_{\mathrm{cut}}$) provide direct constraints on the electron temperature ($kT_{\mathrm{e}}$) and optical depth ($\tau$) within Comptonisation models. These quantities can then be compared with the predictions of pair-regulated scenarios, offering a powerful diagnostic of whether pair processes play a significant role in these systems. In a purely thermal, pair-regulated corona, the high energy cut-off is expected to trace the equilibrium electron temperature set by the balance between heating, Compton cooling, and pair production. In this framework, sources should cluster near the pair thermostat limit in the compactness–temperature plane. However, several AGNs have been observed at significantly lower electron temperatures \citep[e.g.][]{Baloković_2015,10.1093/mnras/stx792,refId0,10.1093/mnras/staa3377,1ursini2016}, suggesting that a purely thermal description may be incomplete.

In addition to this primary hard X-ray continuum, many AGNs exhibit a prominent soft X-ray excess below $\sim$1--2 keV, whose physical origin remains debated. Proposed explanations include thermal Comptonisation in a warm, optically thick medium, relativistic reflection from an ionised accretion disc, or more complex absorption scenarios \citep[e.g.][]{crummy2006,fabian2009broad,walton2013suzaku,liebmann2018x}.
A physically motivated framework that has gained increasing support is the two-corona model (e.g. \citealp{Magdziarz1998,done2012intrinsic,jin2012combined,petrucci2013multiwavelength,Petrucci2018,middei2020,vaia}), in which the X-ray spectrum is produced by two distinct Comptonising regions: a compact, hot corona responsible for the hard X-ray emission, and a warmer, optically thick corona accounting for the soft excess.

Despite decades of observational and theoretical efforts, the physical properties of these coronal components (such as their temperature, optical depth, and geometry) remain key open questions in AGN astrophysics.
At the highest luminosities ($L_{2-10\,\mathrm{keV}} \gtrsim 10^{45}\,\mathrm{erg\,s^{-1}}$) in particular, only a limited number of sources are sufficiently close to allow direct spectral constraints, leaving the coronal properties in this extreme regime still poorly explored.
Radio-quiet quasars provide the cleanest laboratories for studying intrinsic coronal emission, as their X-ray spectra are not contaminated by relativistic jet contributions. 
Systematic studies of the brightest objects at low to intermediate redshift, such as those drawn from the \textit{ROSAT Bright Survey} (RBS; \citealt{Schwope2000}), have revealed a remarkably homogeneous class of objects. 
The \textit{XMM-Newton} analysis of the most luminous RBS quasars by \citet{Krumpe2010}
demonstrated that these sources share smooth soft excesses and hard X-ray continua
consistent with standard coronal emission, together with Fe\,K$\alpha$ lines with typical equivalent widths of
$\mathrm{EW}\simeq100\,\mathrm{eV}$.
Subsequent broadband and multi-epoch studies of single sources, such as RBS~1055, have further shown that variability-based analyses can provide valuable insights into the coupling between the hot corona and distant reprocessing material in these luminous systems \citep{Marinucci2022}. At comparable redshifts, optically selected samples such as the PG quasars
\citep[e.g.,][]{Piconcelli2005}, based on a systematic \textit{XMM-Newton} analysis of
$\sim$40 sources, provide a complementary view, highlighting the intrinsic dispersion
of continuum slopes, absorption features, and soft-excess parametrisations that
are largely suppressed in X-ray-selected samples.

In this context, the quasars PG~1407+265 and RBS~229 represent two of the brightest unlensed, radio-quiet AGNs currently accessible to deep X-ray spectroscopy. 
PG~1407+265 ($z=0.940$; \citealt{McDowell1995}, $R_{\mathrm{L}}\equiv\frac{F_{\nu}(5\mathrm{GHz})}{F_{\nu}(4400\text{\AA})}\simeq3.34$;  \citealt{kellermann1989vla}) is an extremely luminous quasar ($L_{2-10} \approx 9\times10^{45}\,\mathrm{erg\,s^{-1}}$) characterised by a steep X-ray continuum and a strong, variable soft excess \citep{George2000,Reeves2000}. 
A systematic \textit{XMM--Newton} study of PG~1407+265 within the context of the brightest PG quasars \citep{Piconcelli2005} revealed an exceptionally smooth continuum, with weak or absent Fe\,K$\alpha$ emission and no detectable reflection features, placing the source among the most extreme steep-spectrum quasars in the PG sample. 
RBS~229 ($z=0.334$; \citealt{Schwope2000}, $RL\sim1$; \citealt{Krumpe2010}) is one of the most X-ray luminous members of the RBS sample ($L_{2-10 ~keV} \approx 8\times10^{44}\,\mathrm{erg\,s^{-1}}$) and exhibits clear evidence for iron fluorescence and moderate reflection \citep{Krumpe2010}. 

RBS~229 hosts a black hole of mass $M_{\mathrm{BH}}\simeq8.1\times10^{8}\,M_{\odot}$ and radiates at a moderate Eddington ratio, with $L_{\mathrm{bol}}\simeq1.1\times10^{46}\,\mathrm{erg\,s^{-1}}$ and $L_{\mathrm{bol}}/L_{\mathrm{Edd}}\simeq0.1$ \citep{Huang_2023}. In contrast, PG~1407+265 harbours a more massive black hole ($M_{\mathrm{BH}}\simeq1.4\times10^{9}\,M_{\odot}$) accreting at a much higher rate, with $L_{\mathrm{bol}}\simeq1.5\times10^{47}\,\mathrm{erg\,s^{-1}}$ and $L_{\mathrm{bol}}/L_{\mathrm{Edd}}\simeq1.4$ \citep{Punsly_2016}.
Both black hole masses are derived from single-epoch virial estimators, but based on different broad emission lines (H$\alpha$ for RBS~229 and \ion{Mg}{ii} for PG~1407+265). 
{The two sources were selected because they are among the few
extremely luminous radio-quiet quasars with broadband
\textit{XMM--Newton} and \textit{NuSTAR} observations of sufficient
quality to allow a detailed study of their coronal properties.
Their different Eddington ratios provide an additional point of comparison.

In this work, we present a comprehensive broadband X-ray analysis of PG~1407+265 and RBS~229 using recent and archival \textit{XMM-Newton} and \textit{NuSTAR} observations. 
By combining soft and hard X-ray coverage, we have the ability to characterise their spectral variability, reflection properties, and coronal parameters. Throughout this paper, we adopt a flat $\Lambda$CDM cosmology with $H_{0}=70\,\mathrm{km\,s^{-1}\,Mpc^{-1}}$, $\Omega_{\mathrm{M}}=0.3$, and $\Omega_{\Lambda}=0.7$. Spectral fits were carried out with  \texttt{XSPEC} v12.15.1 \citep{Arnaud1996}. All uncertainties are quoted at the 90\% confidence level unless otherwise stated.

\section{Observations and data reduction}

We started by analysing simultaneous \textit{XMM-Newton} and \textit{NuSTAR} observations for each source, obtained over the course of dedicated monitoring campaigns in January 2025. 
For PG~1407+265, the \textit{XMM-Newton} observation was carried out on 18 January 2025, while the corresponding \textit{NuSTAR} coverage was split into two separate pointings carried out on 17 and 22 January 2025, due to scheduling constraints. 
Similarly, RBS~229 was observed by \textit{XMM-Newton} on 31 December 2024, with the associated \textit{NuSTAR} observations divided into three segments obtained on 30 December 2024 and 5 and 16 January 2025 (again as a consequence of scheduling constraints).
This core dataset was further complemented with all available archival \textit{XMM-Newton} and \textit{NuSTAR} observations of PG~1407+265 and RBS~229, resulting in a total of five observations per source. 
A summary of the observations analysed in this work is provided in Table~\ref{tab:obs_all}.

\subsection{XMM-Newton}

The \textit{XMM-Newton} EPIC-pn and EPIC-MOS datasets were processed using SAS v21 \citep{gabriel2004sas}, adopting an iterative procedure to determine  optimal source extraction radii
and time cuts for background flaring, with the goal of maximising the signal-to-noise ratio (S/N), similar to the approach described in \citet{Piconcelli2004}. 
The resulting source extraction radii (reported in Table~\ref{tab:obs_all}) enclose $\sim80-95\%$ of the encircled energy fraction, based on the \textit{XMM-Newton} EPIC PSF calibration.
The background was extracted
from circular regions with radii of 50 arcsec.
Redistribution matrix files were generated using \texttt{rmfgen}, following standard procedures.
Ancillary response files were generated with \texttt{arfgen}, enabling \texttt{applyabsfluxcorr=yes} to improve the cross-calibration between \textit{XMM-Newton} and \textit{NuSTAR}.
Spectra were grouped such that each spectral bin contained at least 30
counts and with a spectral resolution of a factor greater than 3 to avoid oversampling.  

The UV/optical data from the Optical Monitor (OM) onboard \textit{XMM-Newton} were extracted using the standard SAS pipeline (\texttt{omichain}), with the source coordinates retrieved from NED. The filters used in each observation are listed in Table~\ref{tab:obs_all}. The resulting photometric data were processed into spectral files using \texttt{om2pha}.  As noted in the OM calibration documentation\footnote{\url{https://xmmweb.esac.esa.int/docs/documents/CAL-TN-0019.pdf}}, the default photometric errors include only statistical uncertainties; therefore, we included (in quadrature) the additional 3\% systematic uncertainty, as suggested.

\subsection{NuSTAR}

The \textit{NuSTAR} data were reduced with the \texttt{nupipeline} task within NuSTARDAS v2.1.4, using the most recent calibration database (CALDB 20250428 at the time of analysis). The spectra and light curves were extracted with \texttt{nuproducts} for both focal plane modules (FPMA and FPMB), following standard procedures. As with the EPIC-pn data, source and background regions were selected through an S/N optimisation procedure.
The background was extracted
from the circular regions with radii of 100 arcsec, while the optimised source
extraction radii are shown in Table~\ref{tab:obs_all}.
Finally, we applied the optimal spectral binning following the algorithm by \citet{Kaastra2016}, with the additional requirement of a minimum S/N of 3 for each bin. 

\subsection{Swift}

We extracted \textit{Swift}/XRT and \textit{Swift}/UVOT light curves for both sources, providing long-term coverage over several years. 
These data were retrieved from the SSDC archive.\footnote{\url{https://www.ssdc.asi.it}} 
The full list of observations, including ObsID, start time, exposure, and fluxes, is reported in Table~\ref{tab:obs_alphaox_pg_rbs}.

For RBS~229, the same \textit{Swift}/UVOT datasets used to construct the light curves were also employed to extract UV/optical spectra associated with the two \textit{NuSTAR} observations that do not have simultaneous \textit{XMM-Newton} coverage. 
In these cases, the UVOT data provide the necessary UV/optical information contemporaneous with the hard X-ray observations. 
The UVOT spectra were produced using the standard \texttt{uvot2pha} tool, starting from sky images and adopting user-defined source and background regions. 
The set of UVOT filters available for each observation is summarised in Table~\ref{tab:obs_all}.

\section{Variability between short and long timescales}

\begin{figure*}[h!]
\centering
 \includegraphics[width=0.9\linewidth]{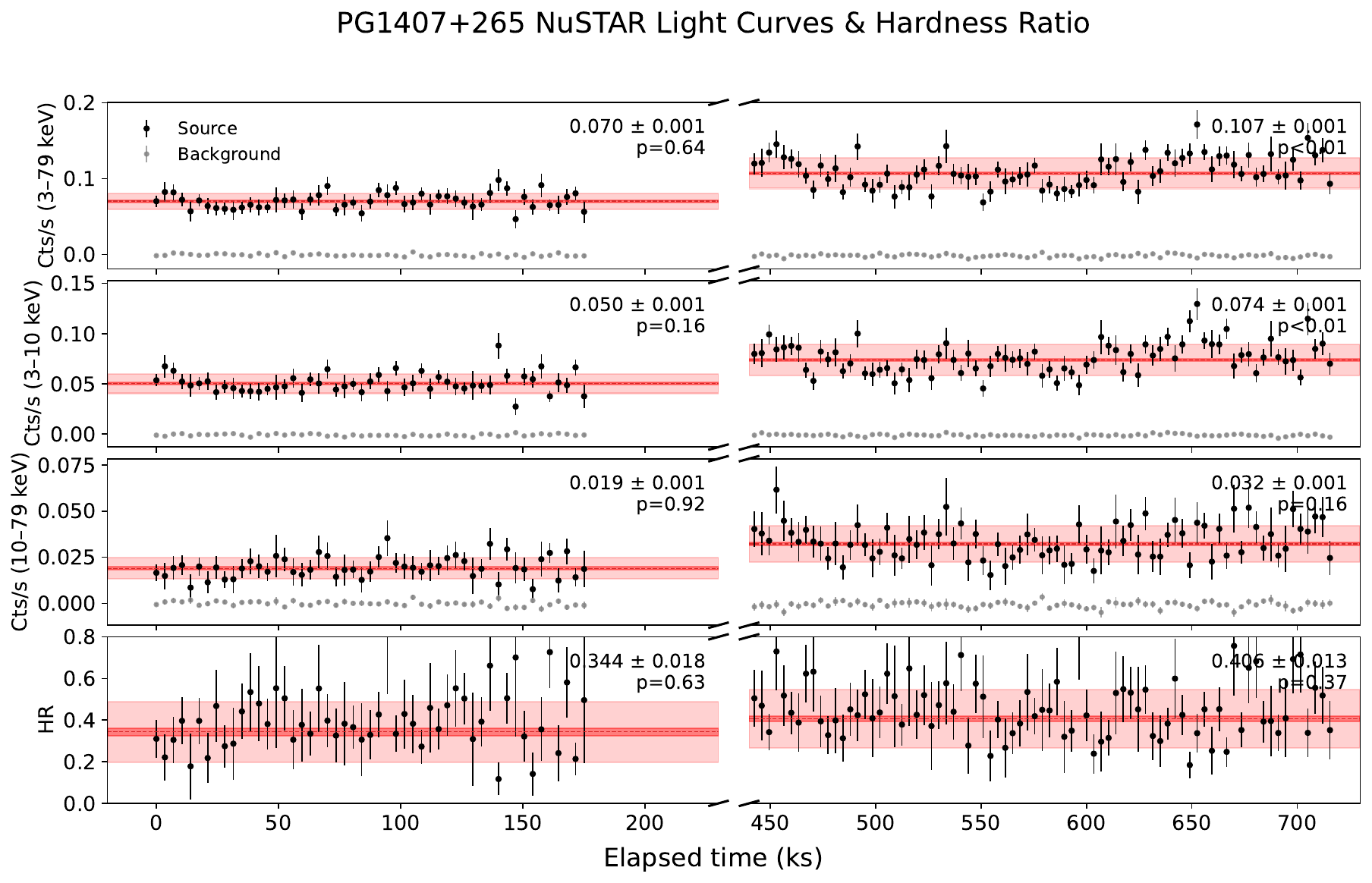}
    \includegraphics[width=0.8\linewidth]{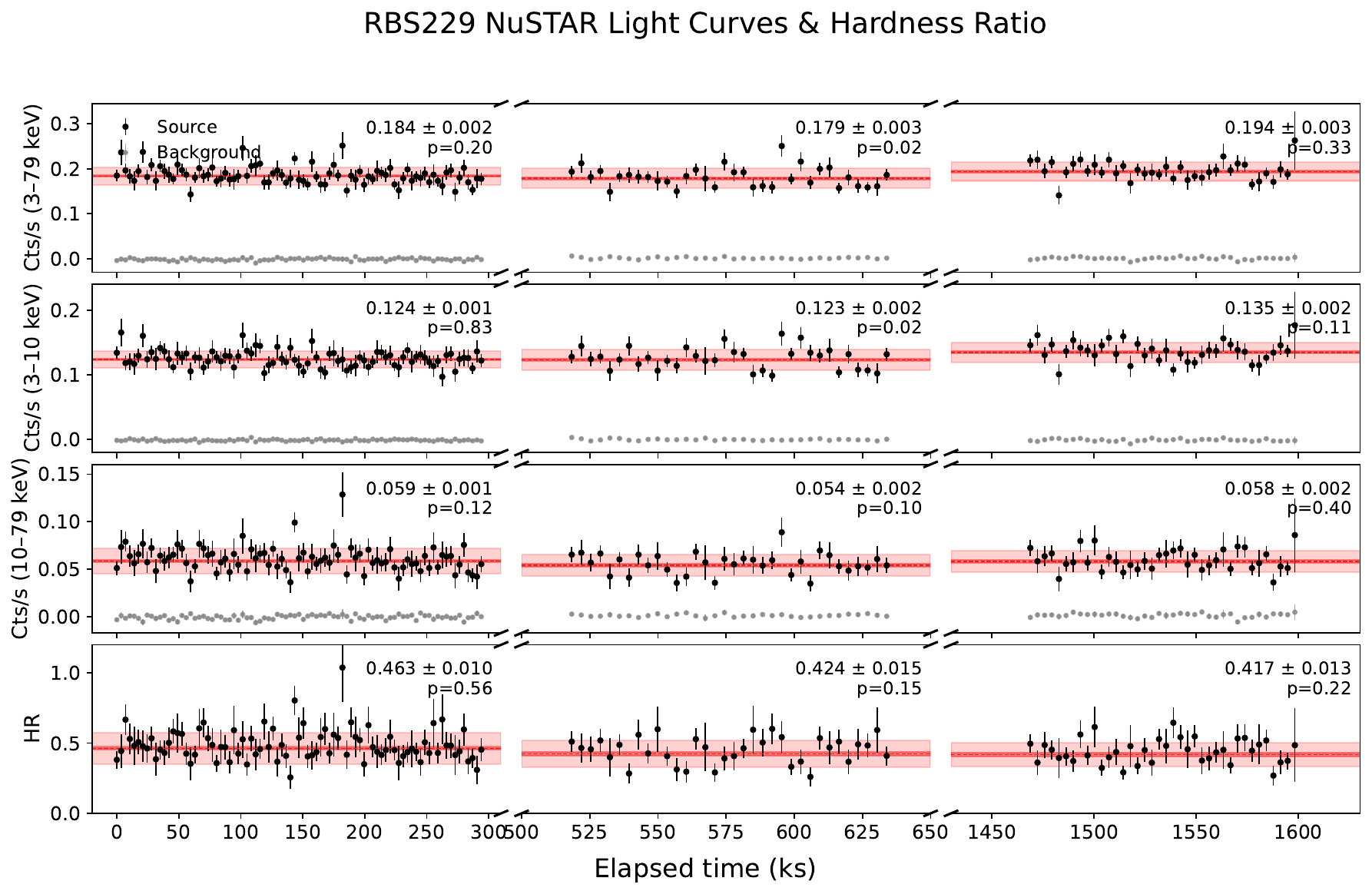}
     \caption{\label{fig:nustar_lc}
Background-subtracted \textit{NuSTAR} FPMA+B light curves of PG~1407+265 (top) and RBS~229 (bottom) in the 3--79~keV, 3--10~keV, and 10--79~keV energy bands.
The bottom panels show the corresponding hardness ratios, defined as $\mathrm{HR} = H/S$, where $H$ and $S$ are the count rates in the 10--79~keV and 3--10~keV bands, respectively. The corresponding background light curves, extracted from the background regions and rescaled consistently with the source extraction regions, are shown for comparison.
The red dashed lines indicate the mean count rates, whose values (with $1\sigma$ uncertainties), together with the $p$-value of the $\chi^2$ test against a constant model, are reported in each panel. The light red shaded regions represent the $\pm1\sigma$ standard deviation of the data points around the mean count rate, while the darker red shaded areas indicate the $1\sigma$ uncertainty on the mean.}
    \label{fig:nustar_total_lc}
\end{figure*}

We investigated the temporal behaviour of the sources over a broad range of
timescales using all available \textit{NuSTAR}, \textit{XMM-Newton}, and
\textit{Swift} observations, focussing initially on short-term variability within
individual observations. The \textit{XMM-Newton} light curves shown in Fig.~\ref{fig:xmm_lc} and the \textit{NuSTAR} light curves shown in Fig.~\ref{fig:nustar_lc} were background-subtracted. For all observations, time intervals affected by background flares were filtered using the same S/N optimisation procedure adopted for the spectral extraction. The only exception is the 2025 observation of PG~1407+265, where the presence of particularly strong background flares led us to adopt the standard SAS high-energy background screening procedure to avoid introducing spurious variability. In all cases, the background light curves were inspected together with the source light curves. No background variations correlated with the source variability discussed below were observed.

The \textit{NuSTAR} light curves, extracted with a time bin of 3500~s and shown
in Fig.~\ref{fig:nustar_lc}, reveal that the \textit{NuSTAR} observations
of PG~1407+265 are characterised by two distinct temporal segments with different
average count-rate levels.
To test for time variability, we fitted the light curves with a constant model and evaluated the resulting $p$-values from a $\chi^2$ test, where the $p$-value represents the probability that the observed variability is due to statistical fluctuations, with $p<0.01$ indicating less than a 1\% chance that the flux is consistent with a constant level.
While the first segment is consistent with a constant flux, the second segment
shows clear intra-observation variability, as indicated by the low p-values in the soft and full energy bands.
This short-term variability is independently confirmed by the contemporaneous
\textit{XMM-Newton} data.
In contrast, the \textit{NuSTAR} hardness ratio remains consistent with being constant throughout the observation ($p = 0.63$ and $p = 0.37$ for the 2025A and 2025B observations, respectively), suggesting that the observed variability is primarily driven by changes in flux rather than spectral shape.

RBS~229 does not exhibit significant short-term variability within
individual \textit{NuSTAR} observations.
None of the individual temporal segments show any statistically significant
intra-observation variability in any of the considered energy bands.
A modest difference in the average count-rate level is observed in the third
temporal segment with respect to the earlier ones; however, this offset is not
associated with enhanced short-term variability.
The hardness ratio remains consistent with being constant across all observations ($p = 0.56$, $0.15$, and $0.22$ for the 2024, 2025A, and 2025B observations, respectively), indicating that the observed differences are primarily driven by changes in flux normalisation, rather than spectral shape.
This behaviour is consistent with the lack of short-term variability observed in the
\textit{XMM-Newton} data (Fig. \ref{fig:xmm_lc}).
For both sources, the stability of the hardness ratio justifies the use of
full-exposure, time-integrated spectra in the subsequent spectral analysis.

Beyond short-term variability, PG~1407+265 also exhibits clear inter-epoch flux changes. In addition to the variability observed between the two \textit{NuSTAR} observations, the flux level measured during the 2001 \textit{XMM-Newton} campaign differs from that observed in the more recent data as shown in Fig.~\ref{fig:xmm_lc}. RBS~229, on the other hand, shows no evidence for variability among its three \textit{NuSTAR} observations. A comparison with the 2015 \textit{XMM-Newton} observation indicates at most modest long-term flux variations.

Characteristic X-ray variability timescales are expected to increase with black-hole mass and decrease with accretion rate \citep{mchardy2006active,gonzalez2012x}. Using the scaling relation of \citet{mchardy2006active}, together with the black-hole masses and bolometric luminosities estimated for our sources, we derive characteristic rest-frame timescales of approximately 4--12 days for PG~1407+265 and 30--42 days for RBS~229. These values are consistent with the weaker short- and long-term variability observed in RBS~229 compared to PG~1407+265.

PG~1407+265, however, does show significant intra-observation variability, most clearly during the 2025 \textit{XMM-Newton} observation. The observed variability corresponds to rest-frame timescales significantly shorter than the characteristic timescales predicted by the \citet{mchardy2006active} relation. This is not necessarily unexpected, since the empirical relation refers to a characteristic (or break) timescale of the X-ray variability power spectrum rather than to a minimum variability timescale. The 2025 \textit{XMM-Newton} exposure corresponds to only a few rest-frame gravitational light-crossing times, making the observed variability physically plausible although suggestive of a very compact emitting region. Significant X-ray variability on both short and long timescales has also been observed in other luminous quasars, indicating that such behaviour is not unexpected in this regime despite the long characteristic timescales predicted by simple scaling relations \citep[e.g.][]{Shemmer_2014,Kammoun2023}.

Finally, we examined the long-term X-ray/UV behaviour using \textit{Swift}. For PG~1407+265, the UVOT $V$-band monitoring reveals clear long-term variability; however, no statistically significant correlation is detected between the UV emission and the soft or hard X-ray flux measured with XRT. In contrast, the soft and hard X-ray fluxes are strongly correlated, with Spearman’s rank coefficients close to unity and $p$-values $\ll 0.01$. Significant long-term variability is detected for RBS~229 in the available \textit{Swift} XRT or UVOT data. More details are given in Appendix \ref{app:swift}.

\begin{figure}[!t]
   \centering
   \includegraphics[width=0.95\columnwidth]{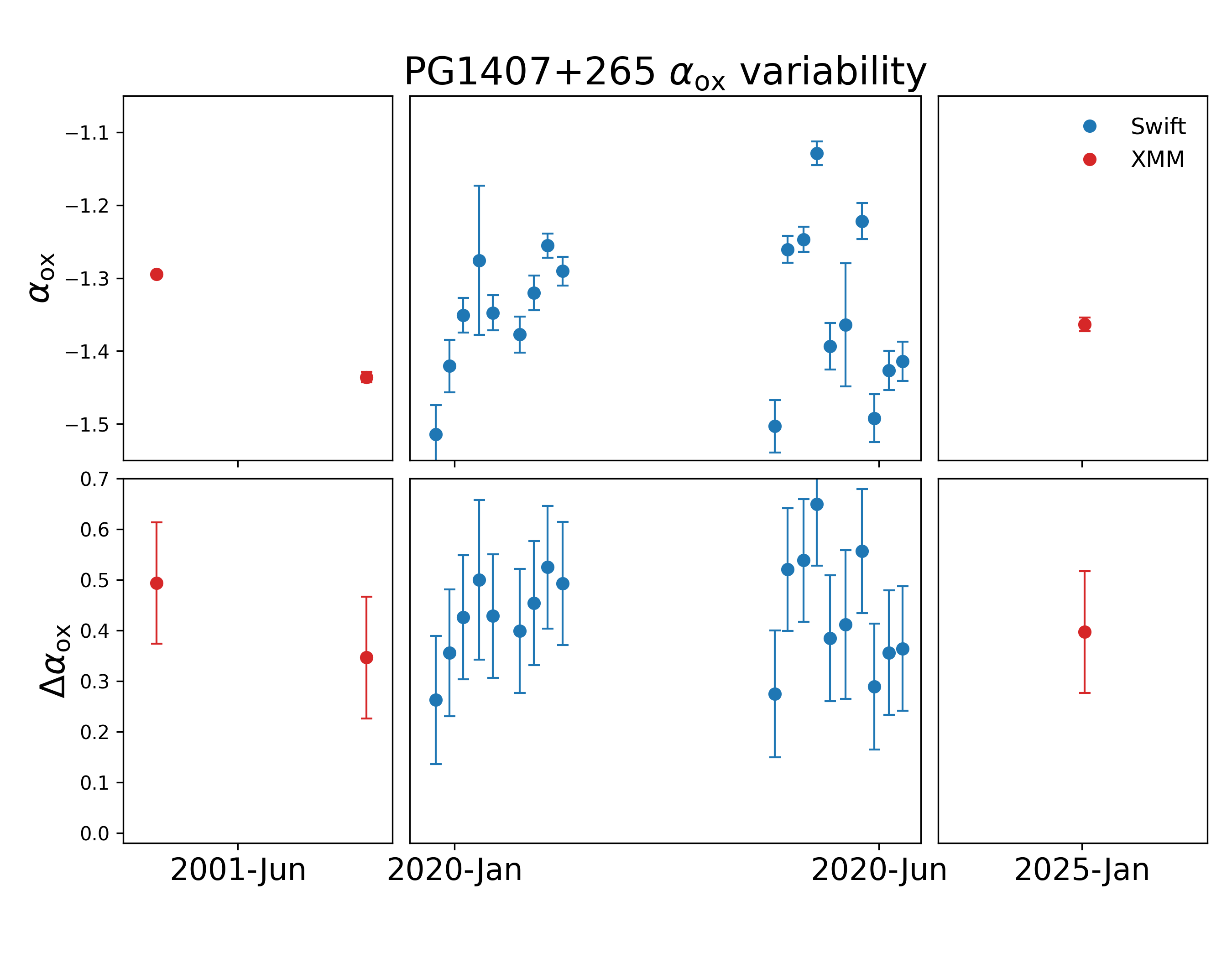}
   \includegraphics[width=0.95\columnwidth]{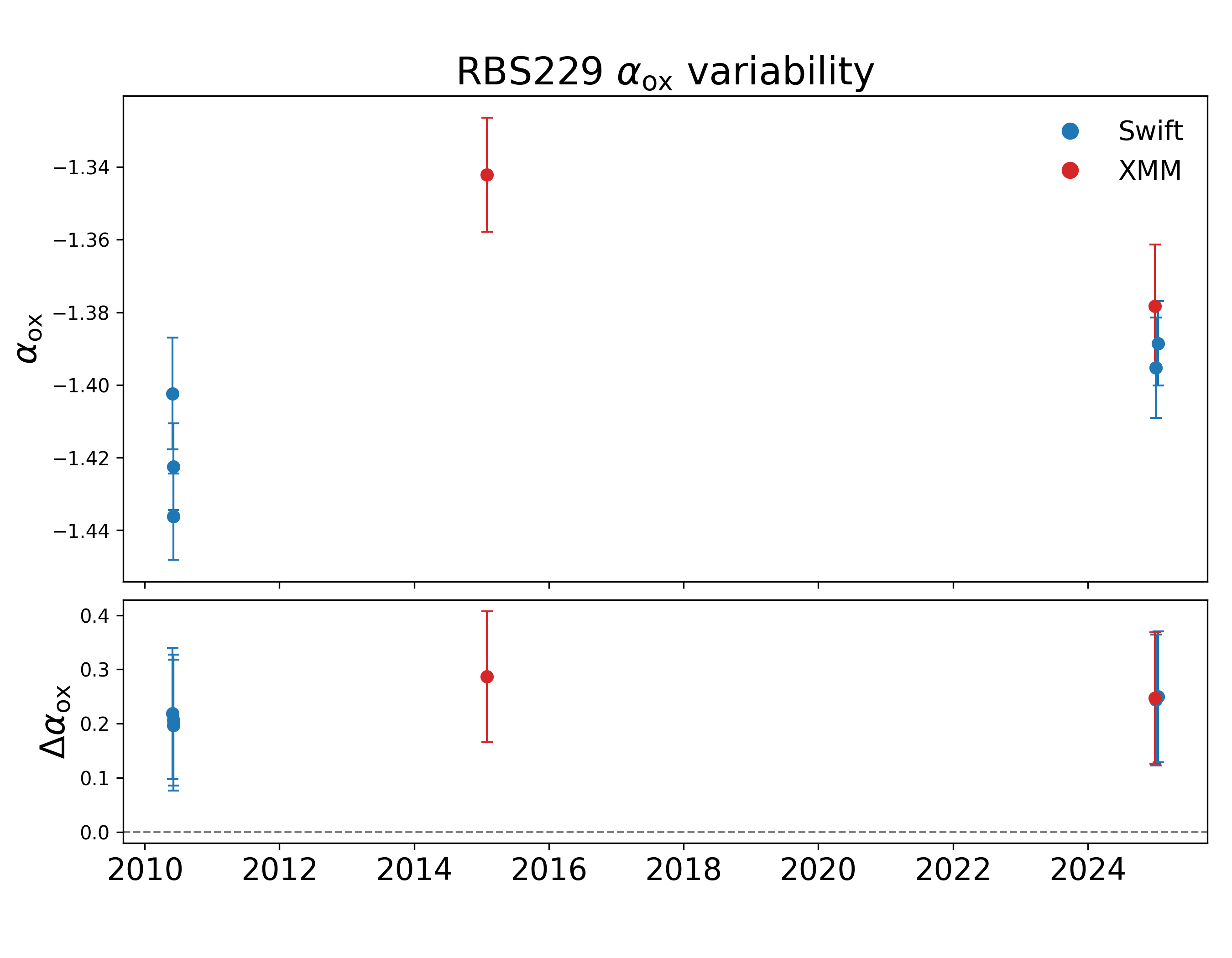}
   \caption{Variability of the $\alpha_{\mathrm{OX}}$ parameter for the two sources. Top: Results for PG~1407+265. Bottom: Same but for RBS~229. In both cases, the top subpanel shows the temporal evolution of $\alpha_{\mathrm{OX}}$, whereas the bottom subpanel displays the corresponding $\Delta\alpha_{\mathrm{OX}}$, defined as the deviation from the value expected from the $\alpha_{\mathrm{OX}}$–UV luminosity relation (see text for details).}
   \label{fig:alphaox_variability}
\end{figure}

For each \textit{Swift} and \textit{XMM--Newton} epoch, we also computed the optical-to-X-ray spectral
index via
\begin{equation}
\alpha_{\rm ox} = 0.3838 \log\!\left(
\frac{f_{2\,\mathrm{keV}}}{f_{2500\,\text{\AA}}}
\right),
\end{equation}
where $f_{2\,\mathrm{keV}}$ is the monochromatic X-ray flux density at
2~keV rest-frame and $f_{2500\,\text{\AA}}$ is the rest-frame UV flux density at
2500~\AA, obtained following the prescriptions by \citet{Gianolli2024} using the available UVOT and OM filters at each epoch.
Expected values were then computed from the $\alpha_{\rm OX}$--$L_{2500}$ relation
of \citet{Vagnetti_2013}, and deviations were quantified as
$\Delta\alpha_{\rm OX} = \alpha_{\rm OX} - \alpha_{\rm OX,exp}$. The resulting $\alpha_{\rm ox}$ and $\Delta\alpha_{\rm ox}$ as a function of time are shown in Fig.~\ref{fig:alphaox_variability}.
Both RBS~229 and PG~1407+265 show significant deviations from the standard
$\alpha_{OX}$--$L_{\mathrm{UV}}$ relation.
In particular, PG~1407+265 displays systematically positive
$\Delta \alpha_{\mathrm{ox}}$ values ($\simeq 0.3$--$0.4$), corresponding to an X-ray
emission approximately $6$--$11$ times higher than expected.
This indicates an X-ray loud behaviour relative to typical quasars.
This result is especially remarkable given that PG~1407+265 was the first discovered weak-line quasar\footnote{Weak-line quasars (WLQs) are quasars characterised by unusually weak or absent high-ionisation broad emission lines in their rest-frame UV spectra (e.g. Ly$\alpha$, \ion{C}{iv}), typically defined by rest-frame equivalent widths significantly lower than those of standard quasars.} \citep{McDowell1995}, a population generally associated with X-ray weak sources \citep[e.g.][]{Cheng2025}.
RBS~229 also shows positive $\Delta \alpha_{\mathrm{ox}}$ values, although of smaller
amplitude, consistent with  moderately enhanced X-ray emission. 
The observed variability in $\alpha_{\mathrm{ox}}$ and $\Delta \alpha_{\mathrm{ox}}$ for both
sources suggests changes in the relative contribution of the corona and
accretion disc emission over time.
The observed variations in $\alpha_{\rm OX}$ appear to be primarily driven by changes in the X-ray emission rather than by variations in the UV continuum. Indeed, as shown in Fig.~\ref{fig:uvot_xrt_lc}, the UV emission remains comparatively stable over the available observations, whereas the soft and hard X-ray fluxes vary significantly.

\section{Spectral analysis}

Our spectral analysis is structured in three successive steps of increasing complexity, designed to progressively characterise the continuum emission, reflection features, and soft-excess component of the spectra. Each step builds upon the previous one and is applied consistently to all available datasets for both sources.

\subsection*{Step 1: 3--10 keV phenomenological continuum}

We started our analysis by fitting all the \textit{XMM-Newton} EPIC pn and MOS spectra of each source in the 3--10~keV energy band with a simple phenomenological model. The model consists of an absorbed redshifted power law
(\textsc{const}$\times$\textsc{tbabs}$\times$\textsc{zpow} in \textsc{xspec}),
intended to provide an adequate description of the primary nuclear continuum and to enable a search for emission features in the residuals.
The Galactic absorption \citep{HI4PI2016} was fixed to
$N_{\mathrm{H}} = 1.15\times10^{20}\,\mathrm{cm^{-2}}$ for PG~1407+265 and
$N_{\mathrm{H}} = 2.65\times10^{20}\,\mathrm{cm^{-2}}$ for RBS~229.
For each observation, the spectral parameters of the MOS data were linked to those of the corresponding pn spectrum, leaving only the cross-calibration constants free to vary. For the pn spectra obtained at different epochs, both the photon index and the normalisation of the power-law component were allowed to vary independently.

\subsection*{Step 2: 3-79 keV hard X-ray modelling}

As a second step, we extend the spectral analysis to the 3--79~keV energy range by jointly fitting all available \textit{XMM-Newton} and \textit{NuSTAR} spectra for each source, adopting a common model across all epochs. Model parameters that do not show significant variability were tied across datasets.
Cross-calibration constants for the \textit{XMM-Newton} spectra were fixed to unity. For each \textit{NuSTAR} epoch, the FPMA cross–calibration constants were tied together and linked to the reference constant of the simultaneous \textit{XMM-Newton} observation, which was left free to vary. The same configuration was adopted for FPMB, resulting in two free cross–calibration constants (one for FPMA and one for FPMB), while the MOS constants were allowed to vary independently. We tested the following spectral models:

\begin{itemize}
  \item Model A: \textsc{const}$\times$\textsc{tbabs}$\times$\textsc{zcutoffpl} \\
  A purely phenomenological description of the primary continuum as an absorbed cut-off power law.

 \item Model B: \textsc{const}$\times$\textsc{tbabs}$\times$(\textsc{zcutoffpl} + \textsc{xillver}). \\
A neutral reflection component is added using \textsc{xillver} \citep{Garcia2013} to reproduce reprocessing features from a plane-parallel slab. The ionisation parameter, inclination, and iron abundance are fixed to $\log(\xi/\mathrm{erg\,cm\,s^{-1}})=0$, $45^\circ$, and solar, respectively. Only the reflected component is included, with the photon index and cut-off energy tied to those of the primary continuum.

  \item Model C: \textsc{const}$\times$\textsc{tbabs}$\times$(\textsc{zcutoffpl} + \textsc{borus}) \\
  Reflection from a toroidal reprocessor is modelled with \textsc{borus }\citep{Baloković_2019}, which self-consistently accounts for both line and continuum emission. The covering factor was fixed to $\mathrm{CF}_{\mathrm{tor}}=0.5$, the iron abundance to solar, and the inclination to $\cos\theta_{\mathrm{inc}}=0.5$.
\end{itemize}

In all models, the normalisations and photon index of the primary continuum components were tied between strictly simultaneous \textit{XMM-Newton} and \textit{NuSTAR} observations. They were also allowed to vary for non-simultaneous datasets.

\subsection*{Step 3: Full broadband modelling}
\label{sec:step3broadbandmodelling}
Finally, we extended the analysis to energies below 3~keV, including the simultaneous \textit{XMM-Newton} OM or \textit{Swift} UVOT data to constrain the UV emission. The following models were tested:

\begin{itemize}
  \item Model D: \\
  \textsc{const}$\times$\textsc{redden}$\times$\textsc{tbabs}(\textsc{nthcomp(W)} + \textsc{nthcomp(H)})

\item Model E: \\
  \textsc{const}$\times$\textsc{redden}$\times$\textsc{tbabs}(\textsc{nthcomp(W)} + \textsc{nthcomp(H)} + \textsc{xillver})
   \item Model F: \\
  \textsc{const}$\times$\textsc{redden}$\times$\textsc{tbabs}(\textsc{nthcomp(W)} + \textsc{nthcomp(H)} + \textsc{borus})
\end{itemize}

Here, \textsc{nthcomp(W)} represents the warm Comptonising corona responsible for the soft excess, while \textsc{nthcomp(H)} models the hot corona producing the hard X-ray emission.
The seed-photon temperature of the warm Comptonisation component was tied across all datasets and treated as a single free parameter. The \textsc{clumin} convolution model was included to compute luminosities directly for both Comptonisation components.
For the reflection components, the spectral parameters were linked to those of the hot Comptonisation component to ensure a consistent illuminating continuum: in particular, the photon index and the electron temperature of the corona ($kT_{\rm e}$) in \textsc{xillver} and \textsc{borus} were tied to those of \textsc{nthcomp(H)}.

To quantify the ratio of the warm and hot components, we introduce the corona strength parameter, denoted here as $S$\footnote{It is referred to as $R$ in \citet{Piconcelli2005}; we adopt the symbol $S$ to avoid confusion with the reflection parameter $R$.}.
In the best fit, all parameters of the \texttt{clumin} component associated with the hot
corona were tied to those of the warm corona, except for a multiplicative constant applied to the warm corona \texttt{clumin}.
This constant directly represents the parameter, $S$, (i.e. the ratio of
the warm-to-hot luminosity in the selected energy band).
The uncertainty on $S$ was determined using the standard error
estimation procedures in \texttt{XSPEC}.
We also derived the Thomson optical depth of the hot and warm Comptonising regions using the phenomenological relation commonly adopted in broadband AGN studies. The best--fit values of the electron temperature and photon index of the \textsc{nthcomp} component were converted into an effective optical depth following the prescription of \citet{Zdziarski1996}, as in \citet{Marinucci2019,Marinucci2022} via

\begin{equation}
\tau \;=\;
\sqrt{
2.25 +
\frac{3}{
\theta\left[(\Gamma+0.5)^2 - 2.25\right]
}
}
-1.5,
\end{equation}
where $\tau$ is the optical depth of the Comptonising plasma, 
$\Gamma$ is the photon index, and 
$\theta = kT_{\rm e}/(m_{\rm e}c^{2})$, 
with $kT_{\rm e}$ the electron temperature, 
$m_{\rm e}$ the electron rest mass, 
and $c$ the speed of light. Uncertainties on $\tau$ were obtained by propagating the 90\% confidence ranges of $(\Gamma, kT_{\rm e})$ through the above relation.
Finally, the \textsc{redden} component accounts for Galactic and intrinsic dust extinction, with reddening values derived from the IRSA Dust Extinction tool based on the dust maps of \citet{Schlegel1998}, recalibrated by \citet{Schlafly2011}.

\subsection{RBS~229}
\label{sec:rbsanalysis}
\subsubsection*{Step 1}
In this first step, the resulting fit provides an overall acceptable description of the 3--10~keV spectra 
($\chi^2 / \mathrm{d.o.f.} = 410/363$). However, an inspection of the residuals reveals localised excesses in the observed 4--6~keV energy range, indicating the presence of additional emission features. 
These deviations are most clearly detected in the 2015 observation, suggesting the presence of narrow emission lines.
We therefore added two redshifted Gaussian components (\textsc{zgauss}) to account for these features, 
with rest-frame centroid energies fixed at 6.4~keV, consistent with neutral iron K$\alpha$ fluorescence, 
and 6.97~keV for Fe\,\textsc{xxvi} Ly$\alpha$.
Allowing the line energies to vary does not significantly improve the fit, and the best-fit values remain 
consistent with these transitions within the uncertainties.
The inclusion of these lines significantly improves the fit, yielding 
$\chi^2 / \mathrm{d.o.f.} = 385/359$. 
The corresponding spectral fits and residuals are shown in Fig.~\ref{fig:ironline_rbs}.

\begin{figure}[h!]
    \centering
\includegraphics[width=0.95\linewidth]{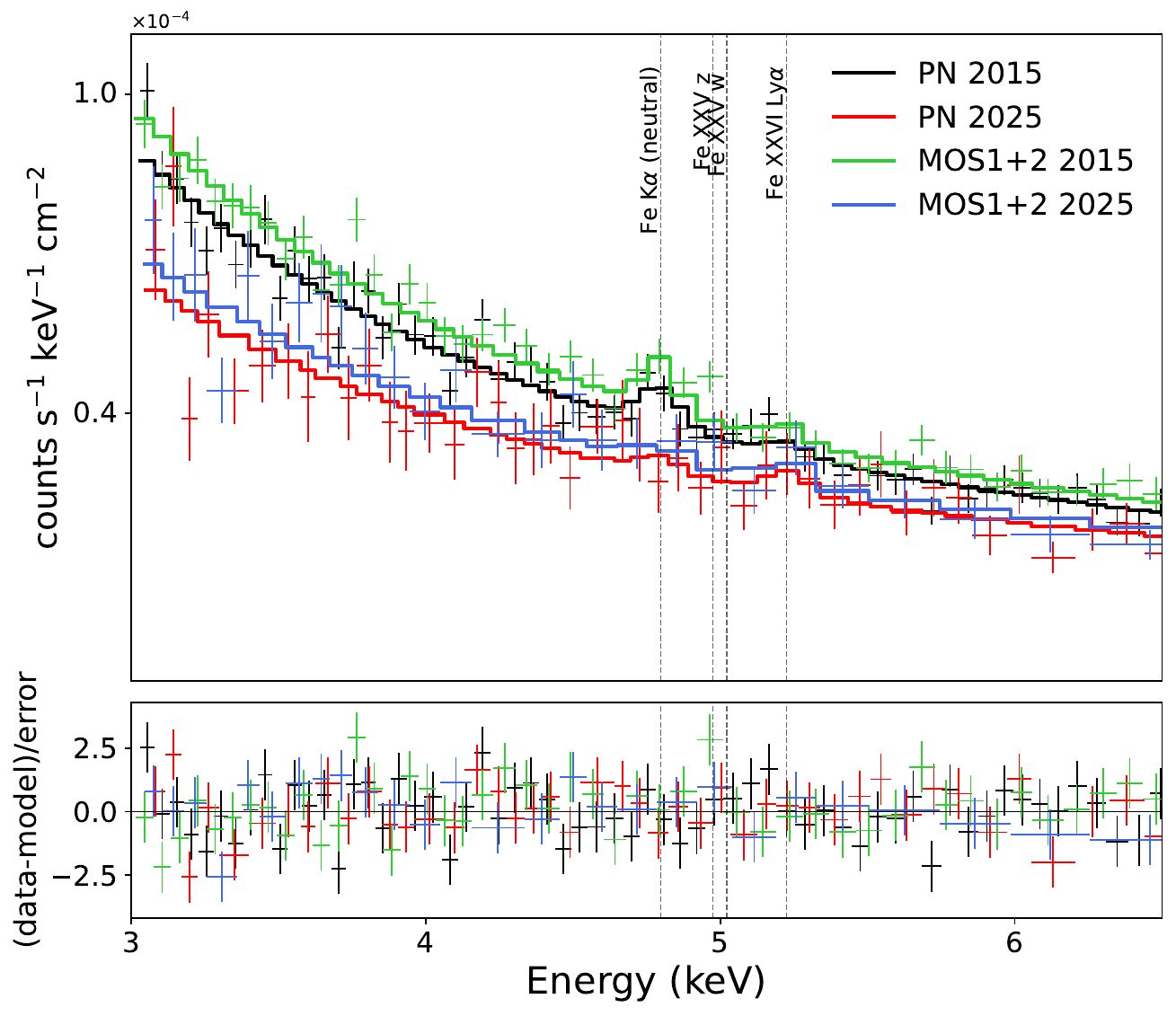}
  \caption{
        3--6.5~keV \textit{XMM-Newton} spectra of RBS~229 for the PN and MOS detectors in different epochs. 
        The solid lines represent the best-fitting power-law models. 
        The dashed vertical lines mark the observed energies of the Fe~K$\alpha$ (neutral), 
        Fe~XXV (w and z lines), and Fe~XXVI~Ly$\alpha$ emission lines, 
        shifted according to the source redshift ($z = 0.334$). 
        The lower panel shows the residuals in units of $(\mathrm{data-model})/\mathrm{error}$.
    }

    \label{fig:ironline_rbs}
\end{figure}

For the 2015 data, both lines are detected, with fluxes of $(1.7\pm0.6)\times10^{-6}$~photons~cm$^{-2}$~s$^{-1}$ (EW $=47\pm16$~eV) for Fe~K$\alpha$ and $(6.4\pm0.3)\times10^{-6}$~photons~cm$^{-2}$~s$^{-1}$ (EW $=20^{+17}_{-1}$~eV) for Fe\,\textsc{xxvi}. For 2025, only the upper limits have been obtained, with fluxes of $<2.2\times10^{-6}$~photons~cm$^{-2}$~s$^{-1}$ (EW $<75$~eV) for Fe~K$\alpha$ and $<2.5\times10^{-6}$~photons~cm$^{-2}$~s$^{-1}$ (EW $<100$~eV) for Fe\,\textsc{xxvi}. In both cases, the measurements are consistent within the statistical uncertainties. All equivalent widths are reported in the rest frame of the source.

\subsubsection*{Step 2}
In the second step, we extended the analysis to the 3--79~keV energy range by including the \textit{NuSTAR} data and performing a combined fit of all available \textit{XMM-Newton} and \textit{NuSTAR} spectra. After binning the spectra to a minimum S/N of 3 per bin, the useful high-energy coverage for RBS~229 extends up to $\sim 40$--$45$ keV
(observed).

As an initial baseline, we model the broadband continuum with a simple \textsc{zcutoffpl} component (model~A), allowing the photon index to vary between epochs to test for spectral slope variability. The resulting best-fit values are consistent within $1\sigma$. The observed inter-epoch variability is instead primarily driven by changes in the continuum normalisation. We therefore tied $\Gamma$ across all datasets, obtaining a common value of $\Gamma = 1.9 \pm 0.1$. Despite providing an acceptable description of the overall continuum, this simple model leaves clear residuals in the Fe~K band and at high energies, characteristic of reflection features such as the iron line complex and the Compton hump.

In agreement with the discussion given in the previous section, the detection of a statistically significant Fe~K$\alpha$ emission line firmly establishes the presence of a reflection component in this source, which must therefore be explicitly included in the spectral modelling. 
The inclusion of the \textsc{xillver} reflection component in model~B significantly improves the fit with respect to model~A, reducing the fit statistic from $\chi^{2}/\mathrm{d.o.f.}=812/721$ to $774/720$. This model yields a photon index of $\Gamma = 1.82 \pm 0.05$ and a high energy cut-off of $E_{\mathrm{cut}} = 70^{+20}_{-40}$~keV. Replacing \textsc{xillver} with \textsc{borus} (model~C) yields a statistically comparable fit, with $\chi^{2}/\mathrm{d.o.f.}=776/720$, and places a lower limit on the cut-off energy of 80~keV, consistent with the value inferred from model~B.
Overall, Models~B and~C offer statistically comparable descriptions of the data, with no significant difference in fit quality. For this reason, we present the results obtained with model~B in Fig.~\ref{fig:modelbvsmodeld} as a representative case.

\begin{figure}[h!]
    \centering
        \begin{subfigure}{0.95\linewidth}
        \centering
        \includegraphics[width=\linewidth]{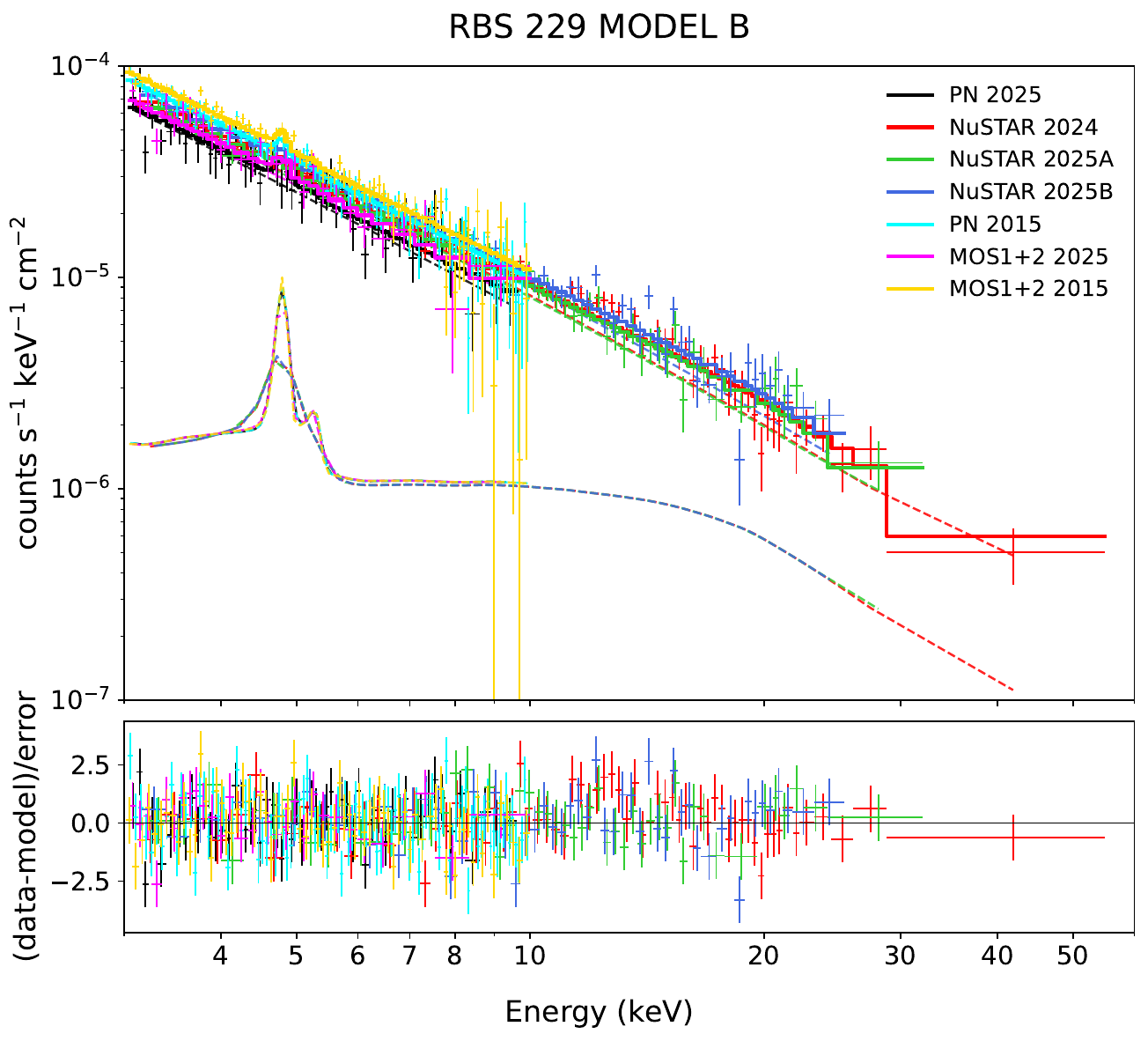}
   \end{subfigure}
\caption{
RBS~229 XMM-Newton and NuSTAR spectra fitted with model~B (\textsc{xillver}). Top: Observed spectra and best-fit model for all instruments. Bottom: residuals, defined as the difference between the data and the model divided by the statistical uncertainty in each bin, $(\mathrm{data}-\mathrm{model})/\mathrm{error}$.
}

    \label{fig:modelbvsmodeld}
\end{figure}

\subsubsection*{Step 3}
\label{sec:step3rbs229}
If we extrapolate the best-fitting hard X-ray continuum to lower energies, a pronounced soft excess emerges below $\sim$3~keV (Fig.~\ref{fig:softexcess_resid}). We therefore extended the modelling to the full broadband range, including the soft X-ray band together with the simultaneous optical/UV data from the \textit{XMM-Newton} OM or the \textit{Swift} UVOT, and fit Models~E and~F to the entire dataset.
In the initial broadband fits, all parameters of both the hot and warm Comptonising components were allowed to vary independently between epochs. While the hot-corona normalisation shows clear variability, the temperature, photon index, and normalisation of the warm Comptonising component are consistent within the uncertainties across all observations. We therefore tied the warm-corona parameters between epochs in the final fits. The apparent differences in the soft-band residuals (Fig.~\ref{fig:softexcess_resid}) are driven by variations in the normalisation of the hot Comptonising component, which modify the underlying continuum level and consequently the apparent strength of the soft excess. We note that no intrinsic variability of the warm corona is required by the data.

Similarly to what was found from the 3-79~keV analysis, both models~E and ~F provide an excellent description of the broadband spectra. However, model~E yields a slightly better fit, with $\chi^{2}/\mathrm{dof} = 1283/1060$, compared to $\chi^{2}/\mathrm{dof} = 1302/1060$ obtained with model~F. For this reason, model~E was adopted as our best-fit model.
In model~E, the hot corona is characterised by a photon index of $\Gamma_{\rm hot} = 1.92 \pm 0.03$ and an electron temperature of $kT_{\rm e,hot} = 13 \pm 3$~keV. The warm Comptonising component is characterised by $\Gamma_{\rm w} = 2.91 \pm 0.07$ and a temperature of $kT_{\rm e,w} = 0.4 \pm 0.1$~keV, while the seed photon temperature is $kT_{\rm bb} = 4.6 \pm 0.6$~eV, consistent with the maximum effective disc temperature predicted by a standard geometrically thin, optically thick accretion disc for the black hole mass and accretion rate inferred for RBS~229 ($\sim$3–13~eV, depending on spin; Eq.~3 of \citealt{LaorandDavis}).
A reflection fraction $R = 0.4 \pm 0.1$ was estimated by computing the ratio between the 20–40~keV luminosities of the primary \textsc{nthcomp} component and of the reflected component (\textsc{xillver}), using the \texttt{clumin} convolution model.
Following the procedure described in Section~\ref{sec:step3broadbandmodelling}, we obtained $\tau_{\rm hot} = 4 \pm 1$ and $\tau_{\rm warm} = 19 \pm 1$. Using the black hole mass reported in the literature and applying an X-ray bolometric correction following \citet{Duras_2020}, we estimated an Eddington ratio of $\lambda_{\rm Edd} = 0.16$, indicating a moderately accreting system. 

All the best-fit parameters are reported in Table~\ref{tab:modelE_modelF}. The corresponding broadband spectral model is shown in Fig.~\ref{fig:broadband}, while a detailed view of the individual spectra for all observations is presented in Fig.~\ref{fig:broadband_plots}.

\subsection{PG~1407+265}
\label{sec:pganalysis}
\subsubsection*{Step 1}
In this first step, the 3-10~keV \textit{XMM-Newton} spectra of PG~1407+265 can be aptly described by a simple absorbed power-law model. The combined pn+MOS spectrum is already well described by the model (\textsc{const}~$\times$~\textsc{tbabs}~$\times$~\textsc{zpow}), yielding $\chi^{2}/\mathrm{dof}=334/315$, and the inclusion of a Gaussian emission line does not improve the fit. We derived 90\% upper limits on the (rest-frame) equivalent width (EW) of a narrow 6.4~keV line of 
$\mathrm{EW} < 39~\mathrm{eV}$ for the 2001A observation,
$\mathrm{EW} < 66~\mathrm{eV}$ for 2001B, and 
$\mathrm{EW} < 130~\mathrm{eV}$ for the 2025 epoch.

\subsubsection*{Step 2}

We extended the spectral analysis of PG~1407+265 to the 3–79~keV band by including the \textit{NuSTAR} data. After binning the spectra to a minimum S/N of 3 per bin, the useful
high-energy coverage extends up to $\sim 30$ keV (observed). As for RBS~229, the photon index was tied across all epochs, since the individual best-fit values are consistent within $1\sigma$.
The spectrum is well described by model~A, yielding $\chi^{2}/\mathrm{d.o.f.} = 551/509$ (Fig.~\ref{fig:modela_pg}). The joint fit gives $\Gamma = 2.1 \pm 0.1$ and constrains the cut-off energy to $E_{\mathrm{cut}} = 80^{+300}_{-40}$~keV. Adding a reflection component (Models~B and~C) does not improve the fit, confirming the absence of any significant reflection contribution, consistent with the iron-line analysis presented above.
\begin{table*}
\caption{Spectral parameters of the two-corona models for
PG\,1407+265 and RBS~229.}
\label{tab:modelE_modelF}
\centering

\begin{tabular}{lccc|cc}
\hline
 & \multicolumn{3}{c|}{\textbf{PG\,1407+265}}
 & \multicolumn{2}{c}{\textbf{RBS\,229}} \\
Parameter & 2001A & 2001B & 2025 & 2015 & 2025 \\
\hline
\multicolumn{6}{c}{Hot Comptonisation and reflection} \\
\hline

$\Gamma_{\mathrm{hot}}$
& \multicolumn{3}{c|}{$2.16\pm0.02$}
& \multicolumn{2}{c}{$1.92\pm0.03$} \\

$kT_{e,\mathrm{hot}}$ [keV]
& \multicolumn{3}{c|}{$>30$}
& \multicolumn{2}{c}{$13\pm3$} \\

$\tau_{\mathrm{hot}}$
& \multicolumn{3}{c|}{$<2$}
& \multicolumn{2}{c}{$4\pm1$} \\

$\log L_{\mathrm{hot}}^{0.5-10}$ [erg/s]
& $46.26\pm0.01$
& $45.93\pm0.01$
& $45.93\pm0.01$
& $45.12\pm0.01$
& $45.01\pm0.01$ \\[4pt]

$R$
& \multicolumn{3}{c|}{$-$}
& \multicolumn{2}{c}{$0.4\pm0.1$} \\

\hline
\multicolumn{6}{c}{Warm Comptonisation} \\
\hline

$\Gamma_{\mathrm{W}}$
& $3.02\pm0.07$
& $3.6\pm0.2$
& $3.28\pm0.09$
& \multicolumn{2}{c}{$2.91\pm0.07$} \\

$kT_{\mathrm{bb}}$ [eV]
& \multicolumn{3}{c|}{$7.6\pm0.6$}
& \multicolumn{2}{c}{$4.6\pm0.6$} \\

$kT_{e,\mathrm{W}}$ [keV]
& $0.8\pm0.1$
& $0.4\pm0.2$*
& $0.4\pm0.2$
& \multicolumn{2}{c}{$0.4\pm0.1$} \\

$\tau_{\mathrm{warm}}$
& $12\pm1$
& $14\pm1$
& $17\pm2$
& \multicolumn{2}{c}{$19\pm1$} \\

$S_{0.5-2}^{\mathrm{W/H}}$
& $0.7\pm0.1$
& $0.2\pm0.1$
& $0.4\pm0.1$
& $0.6\pm0.1$
& $0.7\pm0.1$ \\[4pt]

$S_{0.5-10}^{W/H}$
& $0.37\pm0.06$
& $0.08\pm0.03$
& $0.20\pm0.05$
& $0.26\pm0.04$
& $0.29\pm0.03$ \\[4pt]

$\log L_{\mathrm{warm}}^{0.5-10}$ [erg/s]
& $45.83~\dagger$
& $44.83~\dagger$
& $45.23~\dagger$
& $44.53~\dagger$
& $44.47~\dagger$ \\[4pt]

\hline
\multicolumn{6}{c}{\textit{Total Flux and Luminosity}} \\
\hline

$F_{2-10}$ [$10^{-12}$ erg cm$^{-2}$ s$^{-1}$]
& $1.6\pm0.1$
& $0.77\pm0.02$
& $0.77\pm0.03$
& $2.27\pm0.02$
& $1.73\pm0.02$ \\

$L_{2-10}$ [$10^{45}$ erg s$^{-1}$]
& $8.99\pm0.02$
& $4.00\pm0.05$
& $4.06\pm0.07$
& $0.82\pm0.07$
& $0.63\pm0.08$ \\

$\lambda_{\mathrm{Edd}}$
& $3.35\pm0.01$
& $1.10\pm0.02$
& $1.14\pm0.02$
& $0.22\pm0.02$
& $0.16\pm0.02$ \\

\hline
$\chi^2/\mathrm{dof}$
& \multicolumn{3}{c|}{$1103/981$}
& \multicolumn{2}{c}{$1283/1061$} \\
\hline
\end{tabular}

\tablefoot{ Parameters that were shown to be constant across epochs were tied during the
fit and are therefore reported once, centred across the corresponding
columns. Optical depths were derived as described in the text. The
warm-corona luminosity was computed from
$S^{0.5-10}=L_{\mathrm{warm}}/L_{\mathrm{hot}}$.
The  $\dagger$ symbol next to $L_{\mathrm{warm}}$ indicates that the
warm-corona luminosity was calculated a posteriori using this relation,
that is, derived from $S^{0.5-10}$ and $L_{\mathrm{hot}}$, rather than
obtained as a direct fit parameter. The corresponding uncertainties are
not reported separately because they are implicitly determined through
the uncertainty on $S^{0.5-10}$ evaluated during the fitting procedure.
The 2--10~keV fluxes are observed values, whereas the corresponding
luminosities are intrinsic and corrected for Galactic absorption. The
symbol $^\ast$ indicates that the 2001B warm-corona temperature was tied
to the 2025 value during the fit.
}
\end{table*}
\begin{figure}[h!]
    \centering
    \includegraphics[width=0.95\linewidth]{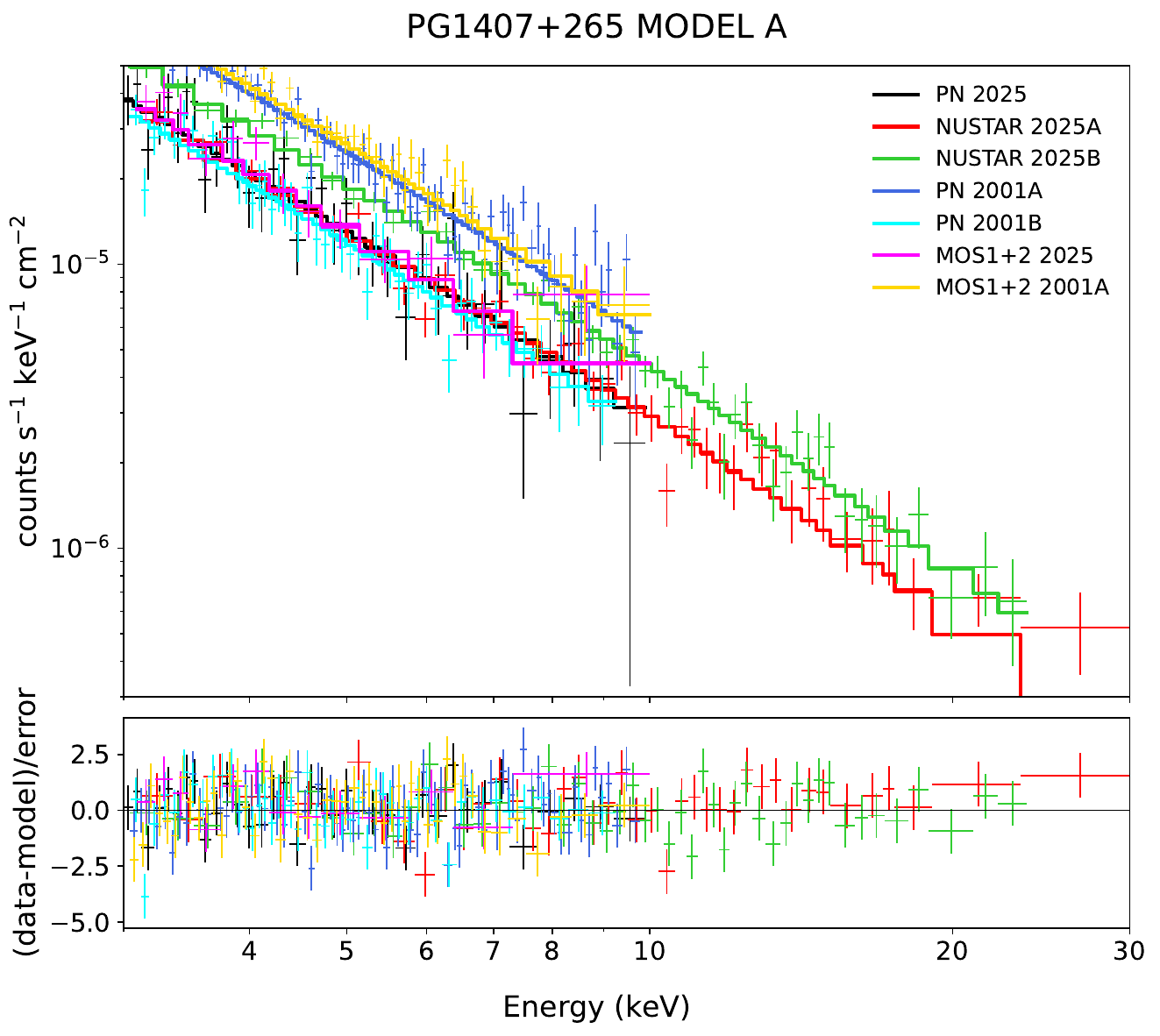}
    \caption{
Best-fit spectrum of PG~1407+265 obtained with model~A (\texttt{zcutoffpl}) in the 3-79 keV band. Plot shows the \textit{XMM-Newton} and \textit{NuSTAR} spectra from different epochs.
Top: Observed spectra and best-fit model. Bottom: Residuals in units of (data–model)/error. This model provides the best statistical fit among those tested, and shows no evidence of reflection features.
}

    \label{fig:modela_pg}
\end{figure}

\begin{figure}[h!]
    
    \begin{subfigure}{1.\linewidth}
        \centering
        \includegraphics[width=\linewidth]{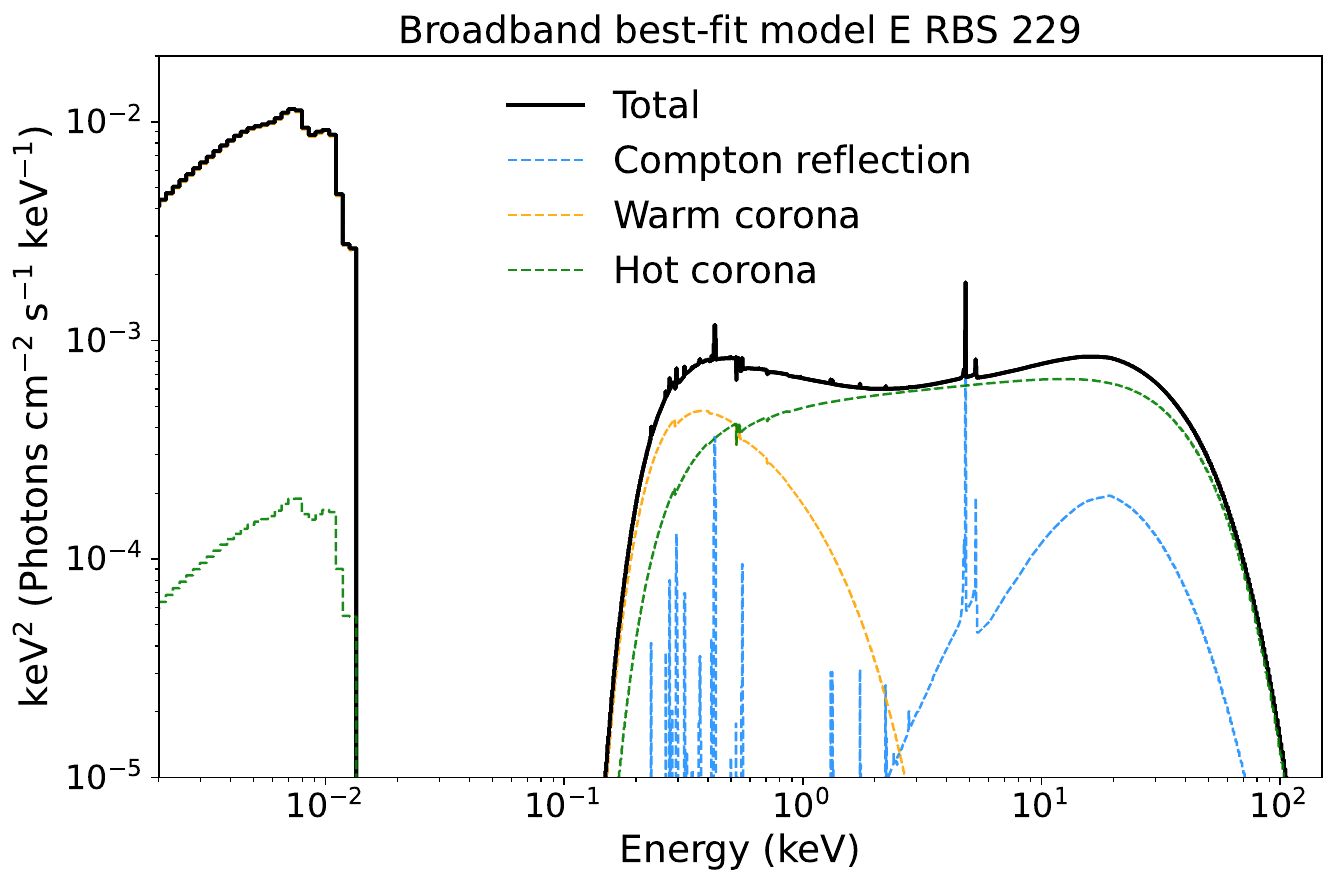}
        
    \end{subfigure}

    \begin{subfigure}{1.\linewidth}
        \centering
        \includegraphics[width=\linewidth]{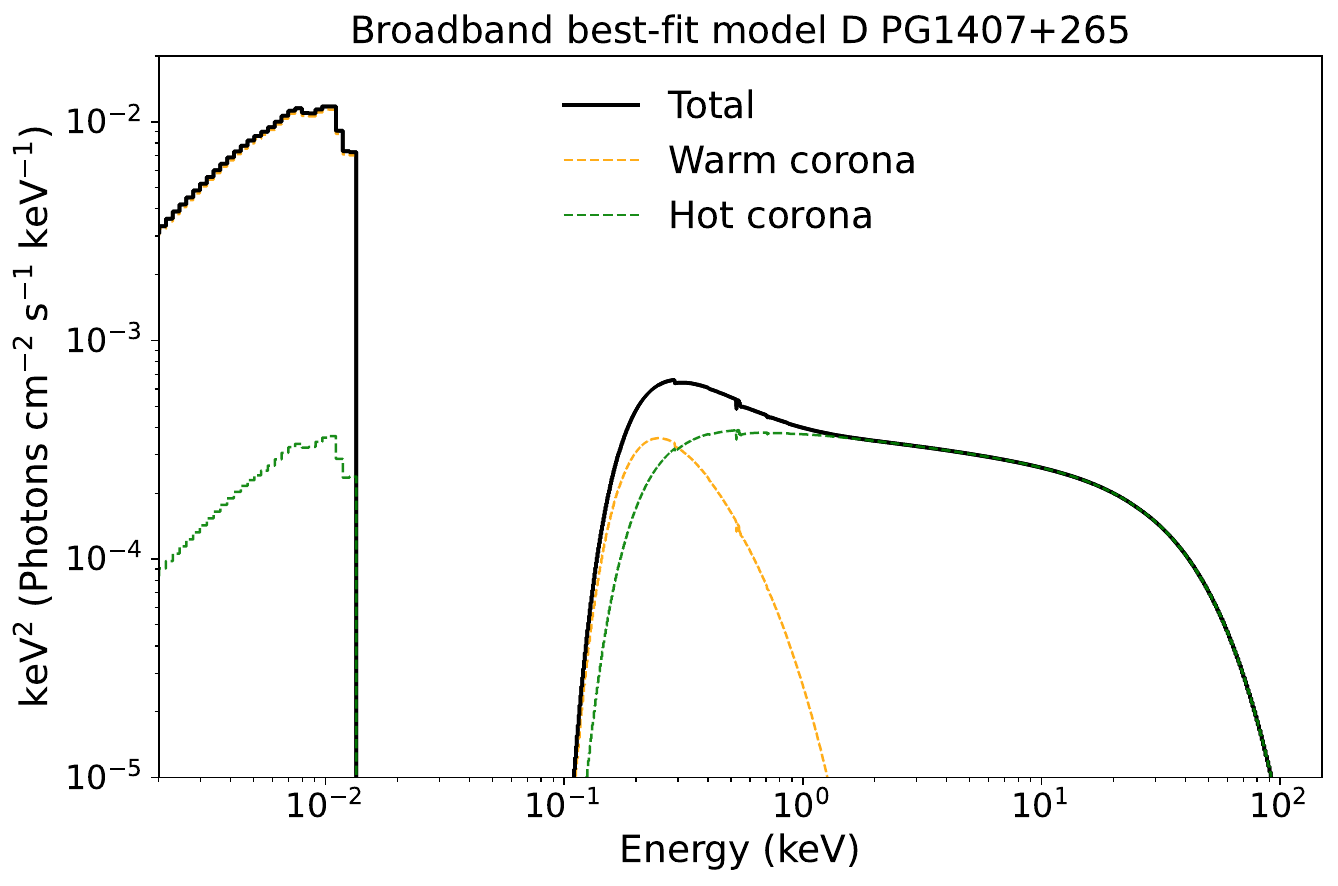}
        
    \end{subfigure}

    \caption{Broadband best-fit spectral models for RBS~229 (top) and PG~1407+265
(bottom). Solid black line shows the total best-fit model, while coloured dashed
curves represent the individual model components. For clarity, only the best-fit models are displayed here; the corresponding spectral data and residuals are shown in the Appendix in Fig. \ref{fig:broadband_plots}.
}
    \label{fig:broadband}
\end{figure}
    
\begin{figure}[h!]
    \centering
 
    \includegraphics[width=0.9\linewidth]{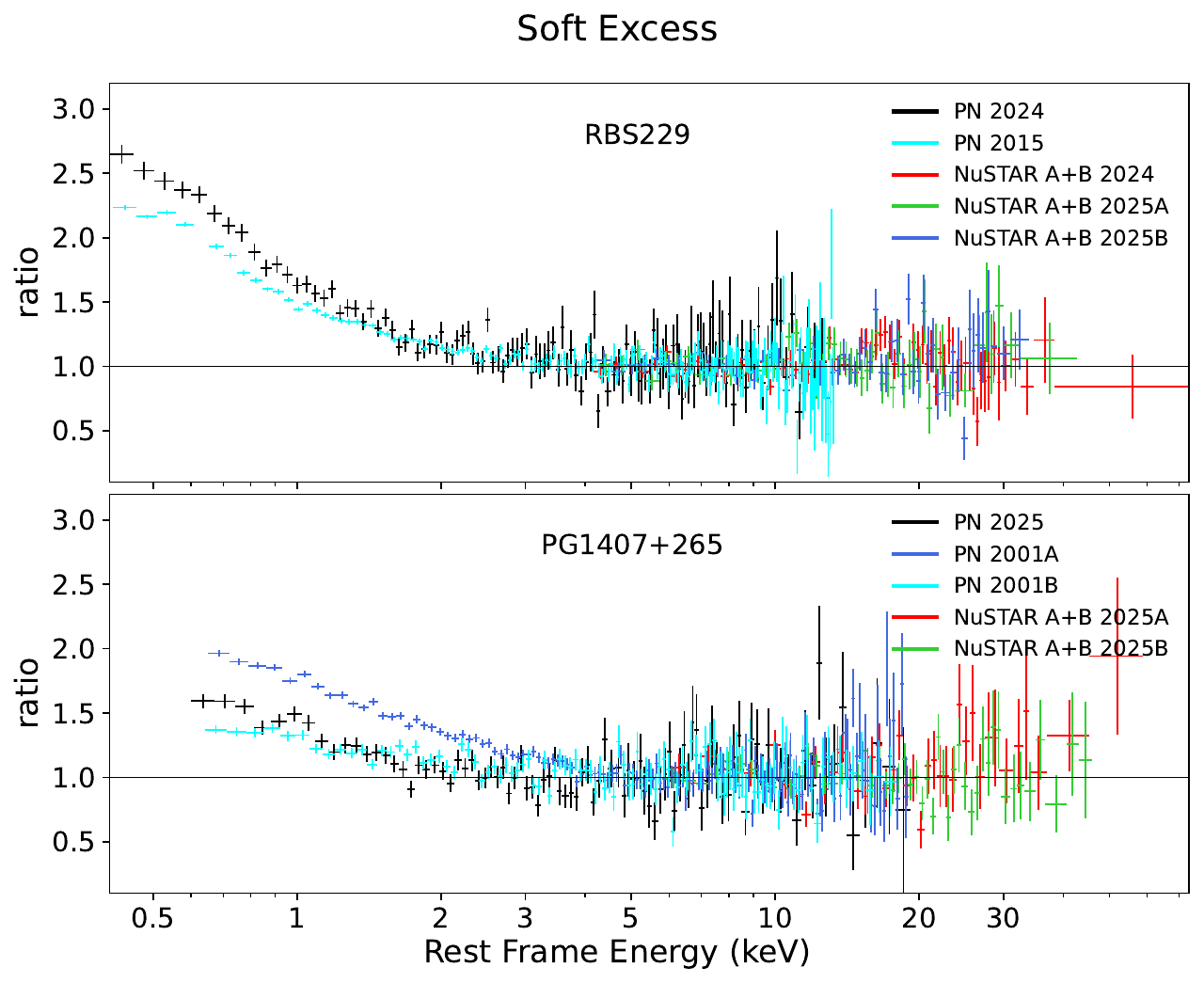}

    \caption{Comparison of the soft--excess components for RBS~229 (top) and PG~1407+265 (bottom). 
For PG~1407+265 the soft excess extends up to $\sim$3~keV in the rest--frame spectrum and shows variability across the different observations. For RBS~229, the soft--excess component is consistent within the uncertainties. For clarity, the MOS spectra are not shown in the plot. The residuals are displayed as ratios.}

 \label{fig:softexcess_resid}
 
\end{figure}
\
    
\subsubsection*{Step 3}
As done for RBS~229, we also extended the spectral analysis of PG~1407+265 to the soft X-ray band by extrapolating the best-fitting hard X-ray continuum to lower energies. In contrast to RBS~229, the soft excess in PG~1407+265 is significantly stronger and more extreme, particularly during the 2001A observation, where it extends up to $\sim$3~keV (see Fig.~\ref{fig:softexcess_resid}), and appears to vary between the 2001A epoch and the 2001B/2025 observations. For this reason, all warm-corona parameters (\textsc{nthcomp(W)}) were allowed to vary independently across the three \textit{XMM--Newton} observations. Since the soft-band spectral shape in 2001B and 2025 appears to be very similar (Fig. \ref{fig:softexcess_resid}), coupled with the fact that the warm-corona temperature cannot be constrained in 2001B, we tied the warm-corona temperatures for these two epochs. Model~D provides a statistically acceptable broadband fit, with $\chi^{2}/\mathrm{dof} = 1103/981$. No additional reflection component is required by the data, as no significant residuals are observed in the Fe~K band or at high energies. Therefore, we adopted model~D as the reference model.

The hot-corona temperature has not been constrained at the 90\% confidence level. The warm corona is clearly detected and well resolved: its photon index and temperature vary across epochs, while the electron temperature is well measured in two observations (2025/2001B and 2001A; see Table~\ref{tab:modelE_modelF} for details). As done for RBS~229, we compared the best-fitting seed photon temperature with the maximum disc temperature predicted by Eq.~3 of \citet{LaorandDavis}, adopting the source parameters. The expected range is $\sim 4$–$21$~eV (from $a=0$ to maximal spin), and the best-fit value $kT_{\rm bb} = 7.6$~eV lies within this interval.

The optical depths of the hot and warm coronae were derived following the same prescription described above in section \ref{sec:step3rbs229}. For the hot component, we obtained an upper limit of $\tau_{\rm hot} < 2.1$. 
For the warm corona, we measured $\tau_{\rm warm,2025} = 17 \pm 2$, $\tau_{\rm warm,2001A} = 12 \pm 1$, and $\tau_{\rm warm,2001B} = 14 \pm 1$.
Adopting the black hole mass from the literature along with the X-ray bolometric correction from \citet{Duras_2020}, we derived Eddington ratios of $\lambda_{\rm Edd} = 1.14$, $3.35$, and $1.10$ for the 2025, 2001A, and 2001B epochs, respectively. These values indicate super-Eddington accretion, particularly during the 2001A observation, which also exhibits the most extreme soft-excess component (Fig.~\ref{fig:softexcess_resid}). All best-fit parameters are reported in Table~\ref{tab:modelE_modelF} and the corresponding broadband spectra are shown in Fig.~\ref{fig:broadband} and Fig.~\ref{fig:broadband_plots}.

\section{Discussion}
\begin{figure*}[t]
  \centering

  \begin{minipage}{0.45\linewidth}
    \centering
    \includegraphics[width=\linewidth]{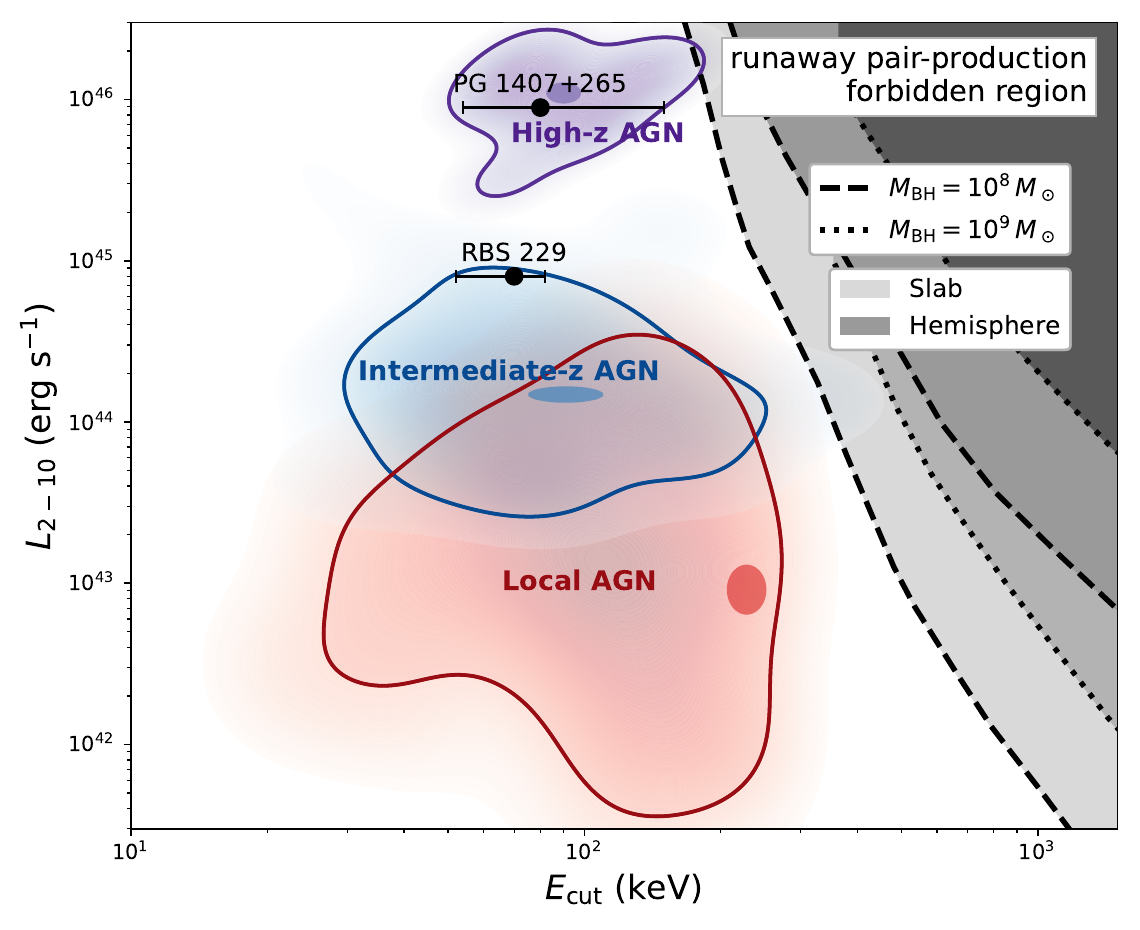}
  \end{minipage}
  \begin{minipage}{0.45\linewidth}
    \centering
    \includegraphics[width=\linewidth]{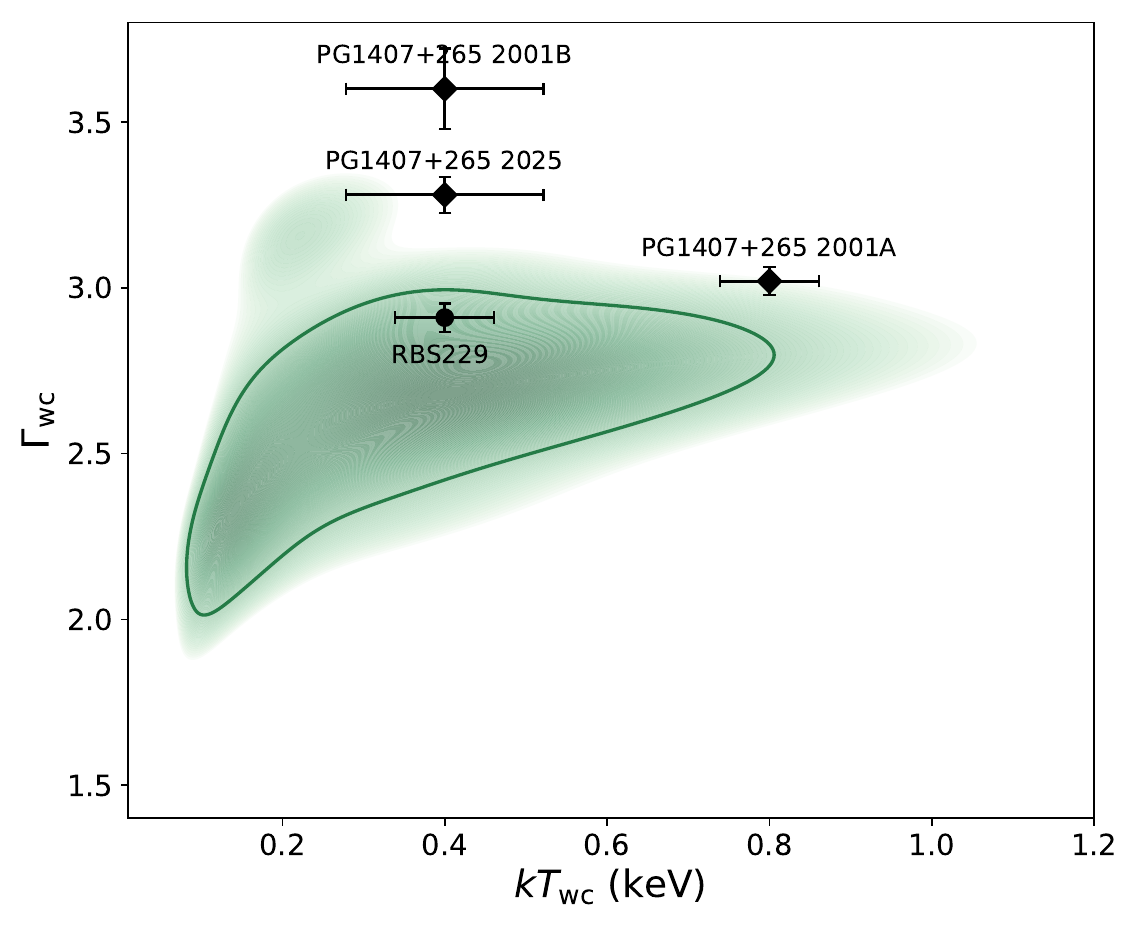}
  \end{minipage}
  \hfill
  \caption{Left: Constraints on the hot corona properties in the $L_{2-10}$--$E_{\rm cut}$ plane. 
Colour-shaded regions represent density distributions of AGN samples at different redshifts: local sources ($z < 0.1$, red), intermediate-redshift sources ($0.1 < z < 0.4$, blue), and high-redshift quasars ($z \gtrsim 1$, purple), adapted from \citet{Bertola_2022} (computed using the nominal $E_{\rm cut}$ values, including lower limits, and not accounting for measurement uncertainties).
For the intermediate-redshift population we also include sources from the SUBWAYS sample \citep{peluso2026supermassive}. 
The contours indicate the iso-density curves at 30\% of the maximum density.
The grey areas mark the runaway pair-production forbidden regions for slab and hemispherical corona geometries. 
Elliptical areas indicate the median values of the different samples reported in \citet{peluso2026supermassive}, where the median values are computed accounting for lower limits through survival analysis techniques (using \textsc{Asurv} and the Kaplan–Meier estimator). 
The black points with error bars show the measurements obtained in this work for PG~1407+265 and RBS~229. Right: Warm corona properties of PG~1407+265 and RBS~229 in the electron temperature--photon index plane, adapted from \citet{Petrucci2018}. 
Coloured regions represent the density distribution of warm-corona parameters observed in AGN samples, shown using the same method adopted in the left panel.}
  
  \label{fig:corona_properties}
\end{figure*}
\subsection{The hot corona}

The hot-corona photon indices derived from the broadband modelling are $\Gamma = 2.16 \pm 0.02$ for PG~1407+265 and $\Gamma = 1.92 \pm 0.03$ for RBS~229. While the latter is still consistent with the typical slopes observed in luminous AGNs \citep[e.g.][]{Piconcelli2005}, PG~1407+265 appears steeper than the average slopes typically observed in local Seyfert samples \citep[e.g.][]{Bianchi2009,Ricci2017} and in intermediate-redshift populations such as SUBWAYS \citep{peluso2026supermassive}. Steep hard X-ray spectra are generally associated with high $\lambda_{\rm Edd}$ (e.g. \citealt{Shemmer2008,Risaliti2009,Laurenti2024,degliAgosti2025}). In this framework, the steeper spectrum of PG~1407+265 relative to RBS~229 is consistent with the variation expected from such $\Gamma$--$\lambda_{\rm Edd}$ trends. The value measured for PG~1407+265 is also in line with
the slopes reported for other high-z and super-Eddington AGNs (e.g. \citealt{Vito2019,Tortosa2023,zappacosta2023}).

In the left panel of Fig.~\ref{fig:corona_properties}, we place PG~1407+265 and RBS~229 in the $E_{\rm cut}$--$L_{2-10}$ plane together with the local, intermediate-redshift (including SUBWAYS), and high-$z$ AGN samples compiled by \cite{peluso2026supermassive}.  The median $E_{\rm cut}$ values of 
the different subsamples, as derived and discussed in that work, are overplotted for reference. Both our sources are shown with their 1$\sigma$ uncertainties on $E_{\rm cut}$, as derived from Step~2 of the 
analysis described in Sects.~\ref{sec:rbsanalysis} and~\ref{sec:pganalysis}. RBS~229 and PG~1407+265 have redshifts that fall within the intermediate-$z$ interval considered by \cite{peluso2026supermassive},
although both lie toward the upper end of that range, particularly PG~1407+265. In terms of luminosity, both sources lie above the median of the intermediate-$z$ population, and PG~1407+265 reaches $L_{2-10}$ values that are more typical of the high-$z$ quasar regime.
RBS~229 shows a measured cut-off energy of $E_{\rm cut}\simeq 70$~keV, consistent with the median values reported for intermediate-$z$ quasars. In addition, despite being an order of magnitude more luminous, PG~1407+265 exhibits a comparable nominal value of $E_{\rm cut}\simeq
80$~keV, albeit with significantly larger uncertainties.

At the population level, \cite{peluso2026supermassive} found that the median $E_{\rm cut}$ decreases with increasing bolometric luminosity and Eddington ratio, while displaying only a weak
dependence on black hole mass. PG~1407+265 is both more luminous and accreting at a higher Eddington ratio than RBS~229, by approximately one order of magnitude in both quantities. Nevertheless, the two sources exhibit comparable high energy cut-offs within the statistical uncertainties. This similarity,
despite the significant difference in luminosity and accretion rate, reflects the intrinsic scatter present in the observed
$E_{\rm cut}$ distributions.

In the $E_{\rm cut}$-$L_{2-10}$ plane of
Fig.~\ref{fig:corona_properties}, both quasars lie safely below the runaway pair-production boundary for plausible assumptions on black hole mass and coronal geometry; therefore, they do not violate the theoretical pair limit. At the same time, their inferred coronal temperatures are significantly lower than the values expected if the corona were operating directly along the pair-regulation
locus \citep[e.g.][]{Fabian2015}. 
This systematic offset, also evident in the broader samples shown in  Fig.~\ref{fig:corona_properties} and discussed by
\cite{peluso2026supermassive} and references therein, suggests that several effects may contribute to shaping the observed high energy cut-offs, including general relativistic effects and hybrid electron distributions \citep[e.g.][]{Tamborra2018,Fabian2017}, as well as geometric and optical-depth effects within the corona \citep[e.g.][]{Middei2019}.

When modelling the primary continuum with a thermal Comptonisation component, we derived for RBS~229 an optical depth $\tau = 4 \pm 1$ and an electron temperature $kT_{\rm e} = 13 \pm 3$~keV, consistent with the median hot-corona value reported for the SUBWAYS sample \citep{peluso2026supermassive}. 
PG~1407+265, on the other hand, requires a hotter and more optically thin plasma, with $kT_{\rm e} > 30$~keV and $\tau < 2$, more akin to the coronal properties commonly inferred in local Seyferts \citep[e.g.][]{Kamraj2022,Pal2024,Serafinelli2024}.
These values of $kT_{\rm e}$ and $\tau$ are consistent with the two-phase disc--corona model in radiative equilibrium \citep{haardt1993x}, in which the balance between Compton cooling and radiative feedback links temperature and optical depth (see their Fig.~1a).

Given the different photon indices of the two quasars, a different combination of temperature and optical depth is naturally expected within thermal Comptonisation models. However, the fact that comparable phenomenological cut-off energies correspond to different inferred temperatures serves as a reminder that the spectra are not simple exponentially cut-off power laws. Care should therefore be exercised when interpreting $E_{\rm cut}$ as a direct proxy for the electron temperature, given the intrinsic degeneracies and systematic uncertainties inherent in Comptonisation models.

Looking at the broadband best-fit SEDs reported in Fig. \ref{fig:broadband}, we can see that the luminosities of the UV bumps appear clearly much more significant than the hard X-ray emission. Actually, the total luminosities of the warm coronae are in the range $\sim (2.3$--$3.0)\times10^{47}$ erg s$^{-1}$ for PG~1407+265 and of the order of $\sim 2.6\times10^{46}$ erg s$^{-1}$ for RBS~229, which are factors of $\sim 5$--$10$ and $\sim 12$ greater than the hot-corona luminosities, respectively. This naturally suggests that only a small part of the accretion power is released in the hot coronae of these two AGNs. In a scenario where the hot coronae are localised in the inner region of the accretion flow, this would also suggest that they ought to be quite compact, with a small radial extension.

\subsection{The warm corona}

The broadband modelling reveals the presence of a prominent soft X-ray excess in both quasars, which is well described by a warm Comptonising corona. In RBS~229, the warm corona is characterised by $\Gamma = 2.91 \pm 0.07$, $kT = 0.4 \pm 0.1$ keV, and $\tau = 19 \pm 1$. Its parameters remain consistent between the 2015 and 2025 observations, including the luminosity. This indicates that the modest variability observed in the soft X-ray band is mainly driven by changes in the hot corona. On the other hand, PG~1407+265 displays a more complex behaviour. While the optical depth remains in the range $\tau \simeq 12{-}17$, the spectral slope and luminosity of the warm component vary significantly across the three epochs. In particular, the warm-corona luminosity changes by nearly an order of magnitude between the two 2001 observations on a timescale of roughly one year; whereas the hot-corona luminosity varies much less ($\sim 0.3$ dex), indicating that the variability is primarily associated with the warm Comptonising component. A relatively stable soft excess, with variability primarily driven by the hot corona, has been observed in several objects \citep[e.g.][]{Middei2018,1ursini2016}, whereas other sources display the opposite behaviour, with the warm Comptonising component dominating the variability, even independently of the hard X-ray continuum \citep[e.g.][]{Mehdipour2011,petrucci2013multiwavelength,Porquet2018}. The contrasting behaviours of RBS~229 and PG~1407+265 therefore place them at opposite ends of this phenomenological range, with the latter extending the soft-excess–driven variability regime to the high-luminosity quasar population.
It is important to note that even in the case of the strongest soft excess considered here (2001A in PG~1407+265), the emission at 2 keV (rest frame) remains dominated by the hot corona. The warm Comptonising component therefore introduces a negligible bias in the derived $\alpha_{\rm OX}$, with $\Delta\alpha_{\rm OX} \sim 0.04$ even in this extreme case. Thus, the observed $\alpha_{\rm OX}$ variability can be interpreted primarily as variability of the hot-corona luminosity, rather than as a direct tracer of changes in the soft-excess component.

In the right panel of Fig.~\ref{fig:corona_properties}, we compare the warm-corona parameters of our sources with those observed in nearby AGNs. In the $\Gamma$--$kT$ plane, RBS~229 falls within the region populated by local Seyferts, as defined by the sample of \citet{Petrucci2018}. Its parameters ($\Gamma = 2.91 \pm 0.07$, $kT = 0.4 \pm 0.1$ keV) are fully consistent with the typical range found in these objects. PG~1407+265 instead occupies more extreme regions of this diagram. The 2001A observation is characterised by a relatively high temperature ($kT \simeq 0.8 \pm 0.1$ keV), while the later epochs favour lower temperatures ($kT \simeq 0.4$ keV) but very steep photon indices, reaching $\Gamma = 3.6 \pm 0.2$ in 2001B and $\Gamma = 3.28 \pm 0.09$ in 2025, placing the source at the upper edge of the parameter space typically explored by warm-corona models. This region of the parameter space agrees with a disc-warm corona structure where part of the disc emission is intrinsic to the disc and not simply due to the reprocessing of the emission coming from the warm corona itself. This addition of disc emission would increase the warm corona cooling and could thus explain its softer spectral shape. The optical depths inferred for the warm coronae are $\tau = 19 \pm 1$ for RBS~229 and $\tau \simeq 12{-}17$ for PG~1407+265, values consistent with those measured in local Seyferts and samples of luminous AGNs (e.g.\ \citealt{Petrucci2018}; \citealt{Palit}; \citealt{peluso2026supermassive}), as well as recent studies of individual luminous quasars such as HE~1029$-$1401 (e.g.\ \citealt{vaia}).

The strength of the soft excess in our sources can be compared with that observed in PG quasars \citep{Piconcelli2005}. For RBS~229, the values are close to the median of the PG sample, with $S_{0.5-10} \simeq 0.3$ and $S_{0.5-2} \simeq 0.6{-}0.7$. PG~1407+265 instead shows a much larger variability, with $S_{0.5-10}$ ranging from $\sim0.08$ to $\sim0.37$ and $S_{0.5-2}$ from $\sim0.2$ to $\sim0.7$ across the three epochs. Nevertheless, all these values remain within the range observed in PG quasars, indicating that the relative contribution of the warm corona is not unusual despite the strong variability observed in this source. 

Based on a large eROSITA AGN sample, \citet{Chen_2025} found that the strength of the soft excess increases with Eddington ratio, while the warm-corona temperature does not show a significant correlation with either $\lambda_{\rm Edd}$ or $M_{\rm BH}$. A consistent picture, in which the physical properties of the warm corona are not solely determined by accretion rate, is also supported by other recent sample studies (e.g.\ \citealt{Waddell2024}; \citealt{Palit}). The comparison between our two quasars is consistent with this view.
Despite the order-of-magnitude difference in luminosity and Eddington ratio between PG~1407+265 and RBS~229, the inferred values of $kT$, $\tau$, and soft-excess strength are broadly similar in the two sources and fall within the ranges observed in other quasars. The main difference is the much steeper warm-corona photon index measured in PG~1407+265. This does not appear to be directly associated with the epoch of highest accretion rate, which is instead characterised by a higher temperature and the highest warm-corona luminosity, while the hot-corona luminosity varies much less. 

Moreover, the variability behaviour of the two quasars is markedly different: while the warm corona in RBS~229 remains stable over a decade, PG~1407+265 shows strong changes in the luminosity of the warm Comptonising component on timescales of about a year. This behaviour is broadly consistent with the two-corona scenario in which the warm Comptonising layer represents the upper atmosphere of the accretion disc and dissipates a significant fraction of the accretion power, so that variations in the accretion flow primarily modulate its luminosity while leaving its thermodynamic properties within a relatively narrow range (e.g. \citealt{petrucci2013multiwavelength,Petrucci2018,KubotaDone2018}). In this context, the warm corona evolution may depend on additional parameters beyond the accretion rate alone, such as the geometry and radial extent of the warm layer, the fraction of power dissipated in the disc atmosphere, and the magnetic field structure (e.g.\ \citealt{Ballantyne2024}; \citealt{Gronkiewicz2023}).

\subsection{Reflection features and the Iwasawa–Taniguchi effect}
Reflection features are weak in both quasars. In PG~1407+265 no reflection component is required by the broadband fits and the Fe~K$\alpha$ line is not detected in any epoch, the most stringent constraint being ${\rm EW} < 39$~eV in the 2001A observation. In contrast, RBS~229 shows evidence for a modest reflection component, with a reflection fraction $R = 0.4 \pm 0.1$ in both epochs and a neutral Fe~K$\alpha$ line detected in 2015 with ${\rm EW} = 47 \pm 16$~eV, while only an upper limit of ${\rm EW} <75$~eV is obtained from the 2025 data.

The weakness of the neutral iron line in both quasars is consistent with the Iwasawa--Taniguchi effect \citep{IwasawaTaniguchi1993}, which predicts a decrease in the Fe~K$\alpha$ EW with increasing X-ray luminosity. Using the relation derived by \citet{Bianchi2007}, the expected equivalent widths are $\simeq25$~eV and $\simeq29$~eV for PG~1407+265 and $\simeq38$~eV and $\simeq40$~eV for RBS~229 at the luminosities measured in the different epochs. The upper limits obtained for PG~1407+265 and the $\simeq47$~eV line detected in RBS~229 are therefore fully consistent with the expectations from the luminosity dependence of the iron-line strength.

The reflection fraction measured in RBS~229 is also consistent with the values typically observed in luminous AGNs in \textit{NuSTAR} surveys, where reflection fractions of $R \simeq 0.3$--$0.5$ are commonly found at $L_{\mathrm{X}} \gtrsim 10^{44}$ erg s$^{-1}$ \citep[e.g.][]{DelMoro2017,Zappacosta2018}. The absence of a detectable reflection in PG~1407+265 is likewise consistent with these trends given its very high luminosity. This behaviour is commonly interpreted as evidence for a decreasing covering factor of the circumnuclear reprocessing material with increasing luminosity.
\section{Conclusions}

We present a broadband X-ray study of the luminous radio--quiet quasars PG~1407+265 and RBS~229 using new simultaneous and archival \textit{XMM--Newton} and \textit{NuSTAR} observations, targeting the coronal structure of AGNs in the high-luminosity regime. Our main results can be summarised as follows:

\begin{itemize}

\item The broadband spectra of both quasars are aptly reproduced by a two-corona scenario consisting of a warm and a hot Comptonising region. This framework provides a self-consistent physical interpretation of both the soft X-ray excess and the hard X-ray continuum, confirming that the two-corona model remains valid even in the regime of luminous and highly accreting radio-quiet quasars.

\item The hard X-ray continua of the two sources show different photon indices, with the steeper spectrum of PG~1407+265 consistent with its much higher accretion rate. Despite the large difference in luminosity and accretion rate between the two quasars, both sources exhibit comparable high energy cut-offs of the order of $\sim$70--80 keV, placing them well below the pair-production limit in the $E_{\rm cut}$--$L$ plane and showing a broad consistency with the trends observed among luminous AGN samples. Comptonisation modelling, however, implies significantly different coronal temperatures, recalling that the phenomenological cut-off energy cannot simply be interpreted as a direct proxy for the electron temperature and reflecting the intrinsic dispersion of coronal properties in luminous AGNs.

\item In RBS~229, the warm corona appears remarkably stable across epochs, while PG~1407+265 displays a strong variability with respect to the warm component. The latter also systematically displays extreme spectral properties of the soft excess, particularly in terms of the very steep warm-corona photon index, extending up to $\sim3$ keV in the rest frame in one epoch. This behaviour is consistent with a scenario in which the warm Comptonising layer represents the upper atmosphere of the accretion disc, with variations in the accretion flow primarily modulating its luminosity while leaving its physical properties within a relatively narrow range.

\item Both quasars show positive deviations from the standard $\alpha_{\rm OX}$--$L_{\rm UV}$ relation, indicating enhanced X-ray emission relative to their UV luminosity. In particular, PG~1407+265 exhibits systematically positive $\Delta \alpha_{\rm OX}$ values, corresponding to an X-ray emission significantly stronger than expected. Reflection features are weak in both sources, with no detectable reflection in PG~1407+265 and only a modest reflection component and Fe~K$\alpha$ line in RBS~229. This is consistent with the decrease in the Fe~K$\alpha$ strength with luminosity, as predicted by the Iwasawa--Taniguchi effect. 
Overall, the two-corona framework successfully describes the broadband emission
of both quasars, despite their different warm-corona behaviour. Extending this
analysis to larger samples with simultaneous broadband observations will be
essential for determining how coronal properties depend on luminosity and
accretion rate.

\end{itemize}

\begin{acknowledgements}
ChatGPT (OpenAI) was used for language editing and code debugging during the preparation of this manuscript. P.O.P. acknowledges financial support from the French ``Action Thematique Phenomenes Extremes et Multimessagers'' from CNRS and from the French spatial agency CNES. E.B acknowledges the support of ``Ricerca Fondamentale 2024'' INAF program (INAF GO grant ``A JWST/MIRI MIRACLE: Mid-IR Activity of Circumnuclear Line Emission'' and mini-grant 1.05.24.07.01. E.P. acknowledges support from the Large Program “DELUX” of the “Ricerca Fondamentale 2024” INAF program. A.L.L. dedicates this work to the memory of his father, who first taught him to look up at the stars. Thank you, Dad.
\end{acknowledgements}

\bibliographystyle{aa}
\bibliography{references}
\clearpage
\onecolumn

\begin{appendix}
\nolinenumbers
\section{Swift monitoring and supplementary material}
\label{app:swift}

This appendix summarises the full set of \textit{Swift} and XMM--Newton OM observations used
in our analysis. Table~\ref{tab:obs_alphaox_pg_rbs} lists the available \textit{Swift}/UVOT and XMM--Newton OM
pointings for PG~1407+265 and RBS~229, together with the corresponding
monochromatic flux densities at 2~keV and 2500~\AA\ and the derived
$\alpha_{\rm ox}$ values.
\begin{figure*}[h!]
    \centering
     \begin{subfigure}{.88\linewidth}
        \centering
        \includegraphics[width=\linewidth]{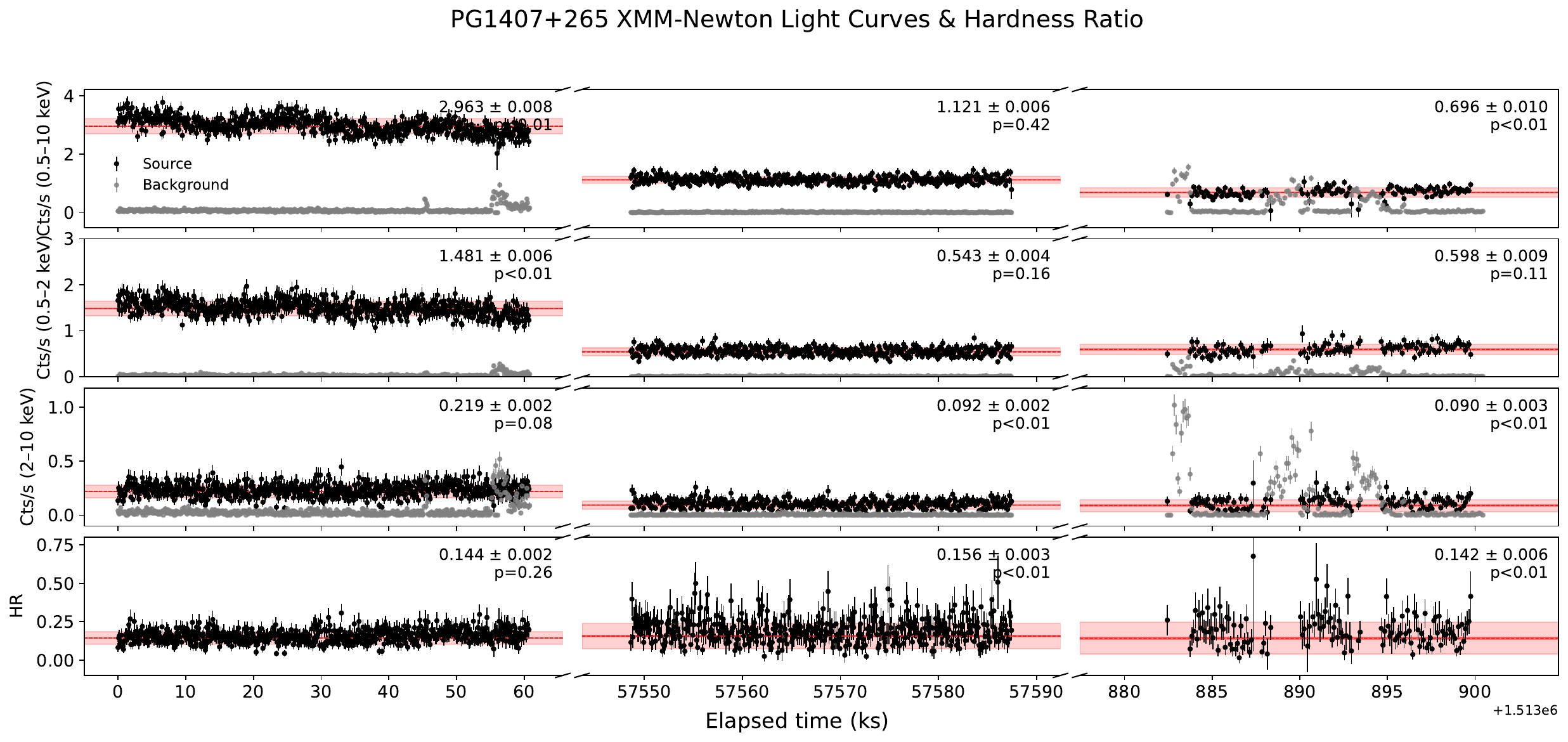}
        
            \end{subfigure}
        \begin{subfigure}{.88\linewidth}
        \centering
        \includegraphics[width=\linewidth]{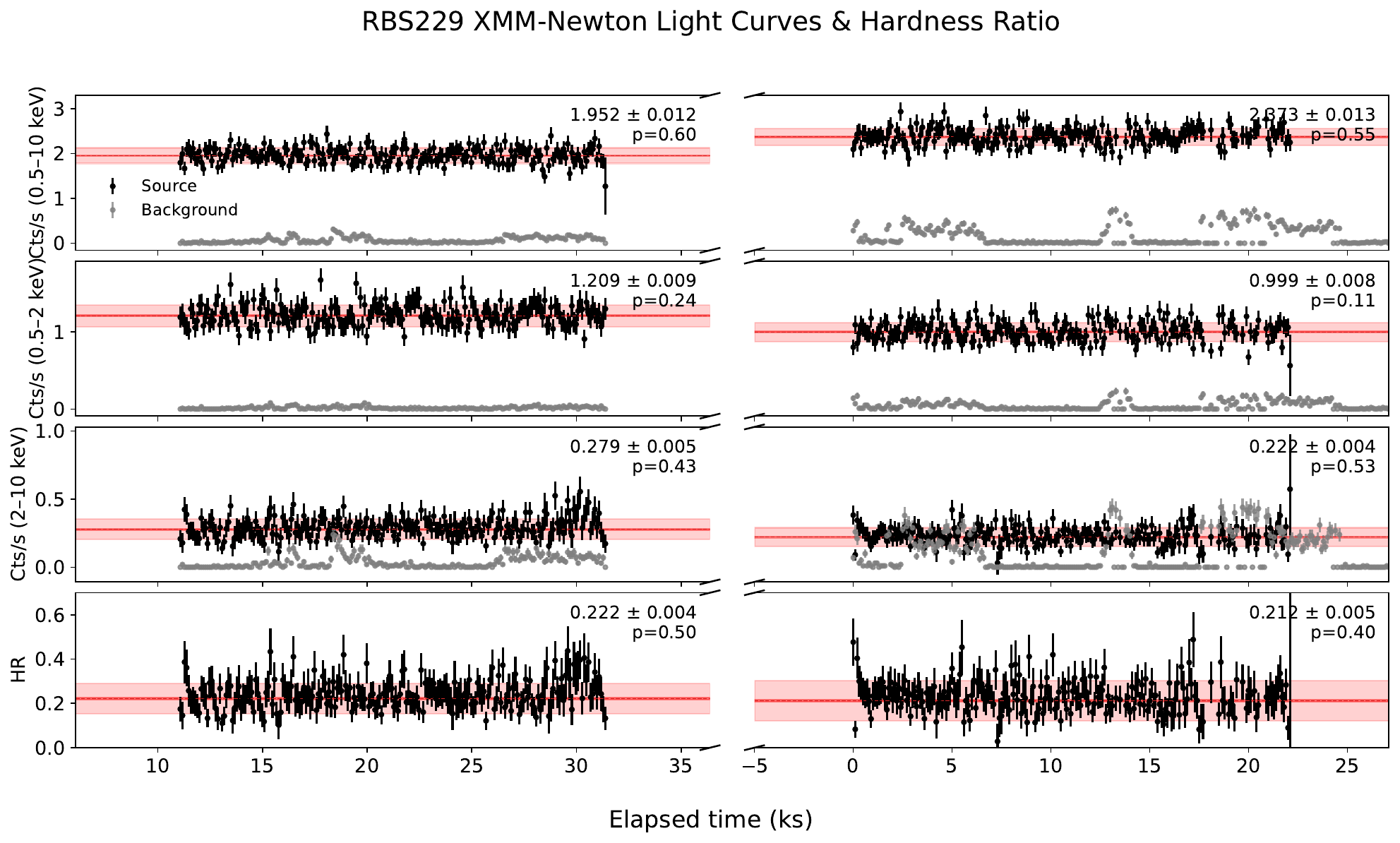}
                    \end{subfigure}
   
    \hfill

   \caption{\label{fig:xmm_lc}
Background-subtracted \textit{XMM-Newton} PN light curves of PG~1407+265 (top) and RBS~229 (bottom) in the 0.5--10~keV, 0.5--2~keV, and 2--10~keV energy bands. The corresponding background light curves, extracted from the background regions and rescaled consistently with the source extraction regions, are shown for comparison.
At the bottom, we also show the corresponding hardness ratios. Red dashed lines indicate the best-fit constant count rates, whose values (with $1\sigma$ uncertainties) are reported in each panel.
Light red shaded regions represent the $\pm1\sigma$ standard deviation of the data points around the mean count rate, while the darker red shaded areas indicate the $1\sigma$ uncertainty on the best-fit constant.
Variability is considered significant when the $p$-value of the $\chi^2$ test against a constant model is $p < 0.01$.
}

\end{figure*}

These measurements are used to reconstruct the long-term UV--X-ray behaviour
of the sources and to characterise their optical-to-X-ray spectral properties
at each epoch.
Figure ~\ref{fig:uvot_xrt_lc} shows the corresponding \textit{Swift}/XRT
and UVOT light curves. For each source, the panels display:
(i) the UV flux densities in all available filters,
(ii) the soft and hard X-ray light curves, and
(iii) the hardness ratio as a function of time.
The monochromatic flux densities shown in the light curves and used for
the computation of $\alpha_{\rm ox}$ were derived as follows.
For each observation, the X-ray spectra extracted from Swift/XRT and the optical/UV photometric points from UVOT were modelled within
\textsc{XSPEC} using a \texttt{zpowerlaw} component, with the redshift
fixed to the source value and Galactic extinction accounted for by
including the \texttt{redden} component.
For the optical/UV data, the photon index was fixed to $\Gamma = 0.5$,
while for the X-ray spectra it was left free to vary during the fit.
The power-law normalisation was set using the \texttt{xset POW\_EMIN} and \texttt{POW\_EMAX} commands, fixing the energy to 2~keV for the X-ray spectra and to the effective wavelength of each UVOT filter for the
optical/UV data to directly obtain the monochromatic flux
density at the corresponding energy or wavelength. When multiple UVOT filters were available for a given epoch, the
rest-frame flux density at 2500~\AA\ was obtained through log--log
interpolation; otherwise, a power-law extrapolation was performed
assuming the fixed spectral slope. 
The optical-to-X-ray spectral index was then computed as $\alpha_{\rm ox} = 0.3838 \, \log \left(\frac{f_{\nu}(2~{\rm keV})}{f_{\nu}(2500~\text{\AA})}\right),$
with uncertainties propagated assuming independent errors on the X-ray
and optical/UV flux densities.

\begin{table*}[b!]
\centering
\caption{Observation data and derived optical-to-X-ray quantities for
PG~1407+265 and RBS~229.}
\label{tab:obs_alphaox_pg_rbs}

\begin{tabular}{lcccccc}
\hline
ObsID & Start date & Exp. & Filter(s) &
$f_{2\,\mathrm{keV}}$ & $f_{2500\,\text{\AA}}$ & $\alpha_{\rm ox}$ \\
 & (UTC) & (ks) & &
(erg s$^{-1}$ cm$^{-2}$ Hz$^{-1}$) &
(erg s$^{-1}$ cm$^{-2}$ Hz$^{-1}$) \\
\hline
\multicolumn{7}{c}{\textbf{PG~1407+265}} \\
\hline
00012832001 & 2020-01-08 & 0.04 & V & $2.68\times10^{-30}$ & $1.96\times10^{-26}$ & $-1.48$ \\
00012832002 & 2020-01-13 & 0.8 & V & $4.62\times10^{-30}$ & $1.93\times10^{-26}$ & $-1.39$ \\
00012832003 & 2020-01-18 & 0.6 & V & $7.13\times10^{-30}$ & $1.96\times10^{-26}$ & $-1.32$ \\
00012832004 & 2020-01-23 & 0.9& V & $1.10\times10^{-29}$ & $1.92\times10^{-26}$ & $-1.24$ \\
00012832005 & 2020-01-28 & 1.1 & V & $7.16\times10^{-30}$ & $1.93\times10^{-26}$ & $-1.32$ \\
00012832007 & 2020-02-07 & 0.8 & V & $6.02\times10^{-30}$ & $1.94\times10^{-26}$ & $-1.35$ \\
00012832008 & 2020-02-12 & 1 & V & $8.23\times10^{-30}$ & $1.88\times10^{-26}$ & $-1.29$ \\
00012832009 & 2020-02-17 & 0.9 & V & $1.34\times10^{-29}$ & $2.08\times10^{-26}$ & $-1.22$ \\
00012832010 & 2020-02-22 & 1.0 & V & $1.14\times10^{-29}$ & $2.17\times10^{-26}$ & $-1.26$ \\
00012832011 & 2020-05-08 & 0.9& V & $2.92\times10^{-30}$ & $2.00\times10^{-26}$ & $-1.47$ \\
00012832012 & 2020-05-13 & 0.9 & V & $1.31\times10^{-29}$ & $2.10\times10^{-26}$ & $-1.23$ \\
00012832013 & 2020-05-18 & 0.9 & V & $1.53\times10^{-29}$ & $2.26\times10^{-26}$ & $-1.22$ \\
00012832014 & 2020-05-23 & 0.9 & V & $2.76\times10^{-29}$ & $2.00\times10^{-26}$ & $-1.09$ \\
00012832015 & 2020-05-28 & 0.9 & V & $5.63\times10^{-30}$ & $2.00\times10^{-26}$ & $-1.36$ \\
00012832016 & 2020-06-03 & 0.9 & V & $6.43\times10^{-30}$ & $1.91\times10^{-26}$ & $-1.33$ \\
00012832017 & 2020-06-08 & 0.1 & V & $1.59\times10^{-29}$ & $2.02\times10^{-26}$ & $-1.19$ \\
00012832018 & 2020-06-13 & 1 & V & $3.30\times10^{-30}$ & $2.11\times10^{-26}$ & $-1.46$ \\
00012832019 & 2020-06-18 & 1 & V & $4.98\times10^{-30}$ & $2.16\times10^{-26}$ & $-1.39$ \\
00012832020 & 2020-06-23 & 1 & V & $4.98\times10^{-30}$ & $2.00\times10^{-26}$ & $-1.38$ \\
\hline
0092850101 (XMM) & 2001-01-23 & 62 & UVW2 & $9.90\times10^{-30}$ & $2.37\times10^{-26}$ & $-1.30$ \\
0092850501 (XMM) & 2001-12-22 & 39 & UVW2 & $4.70\times10^{-30}$ & $2.15\times10^{-26}$ & $-1.41$ \\
0935790201 (XMM) & 2025-01-18 & 15 & UVM2,UVW1 & $5.00\times10^{-30}$ & $1.79\times10^{-26}$ & $-1.36$ \\
\hline
\multicolumn{7}{c}{\textbf{RBS~229}} \\
\hline
00089873001 & 2025-01-05 & 2 & U & $4.20\times10^{-30}$ & $1.81\times10^{-26}$ & $-1.39$ \\
00089873003 & 2025-01-17 & 1 & U & $4.30\times10^{-30}$ & $1.78\times10^{-26}$ & $-1.39$ \\
00040744001 & 2010-05-30 & 2& UVW1 & $2.97\times10^{-30}$ & $1.34\times10^{-26}$ & $-1.40$ \\
00040744002 & 2010-06-04 & 4 & U,UVW1 & $2.99\times10^{-30}$ & $1.52\times10^{-26}$ & $-1.42$ \\
00040744003 & 2010-06-06 &4 & UVW2 & $2.98\times10^{-30}$ & $1.65\times10^{-26}$ & $-1.44$ \\
\hline
0744450301 (XMM) & 2015-01-29 & 137 & UVW2,UVM2,UVW1,U & $4.83\times10^{-30}$ & $1.52\times10^{-26}$ & $-1.34$ \\
0935790101 (XMM) & 2024-12-31 & 20 & UVM2,UVW1 & $3.70\times10^{-30}$ & $1.44\times10^{-26}$ & $-1.38$ \\
\hline
\end{tabular}
\tablefoot{
The columns list the observation identifier, start date, exposure
time, available filters, monochromatic flux densities at 2~keV and
2500~\AA, and the corresponding $\alpha_{\rm ox}$ value. Exposure
times for the \textit{Swift}/UVOT and \textit{XMM-Newton} observations
are given in kiloseconds.
}
\end{table*}

\begin{figure*}[h!]
    \centering

    \begin{subfigure}[t]{0.9\textwidth}
        \centering
        \includegraphics[width=\textwidth]{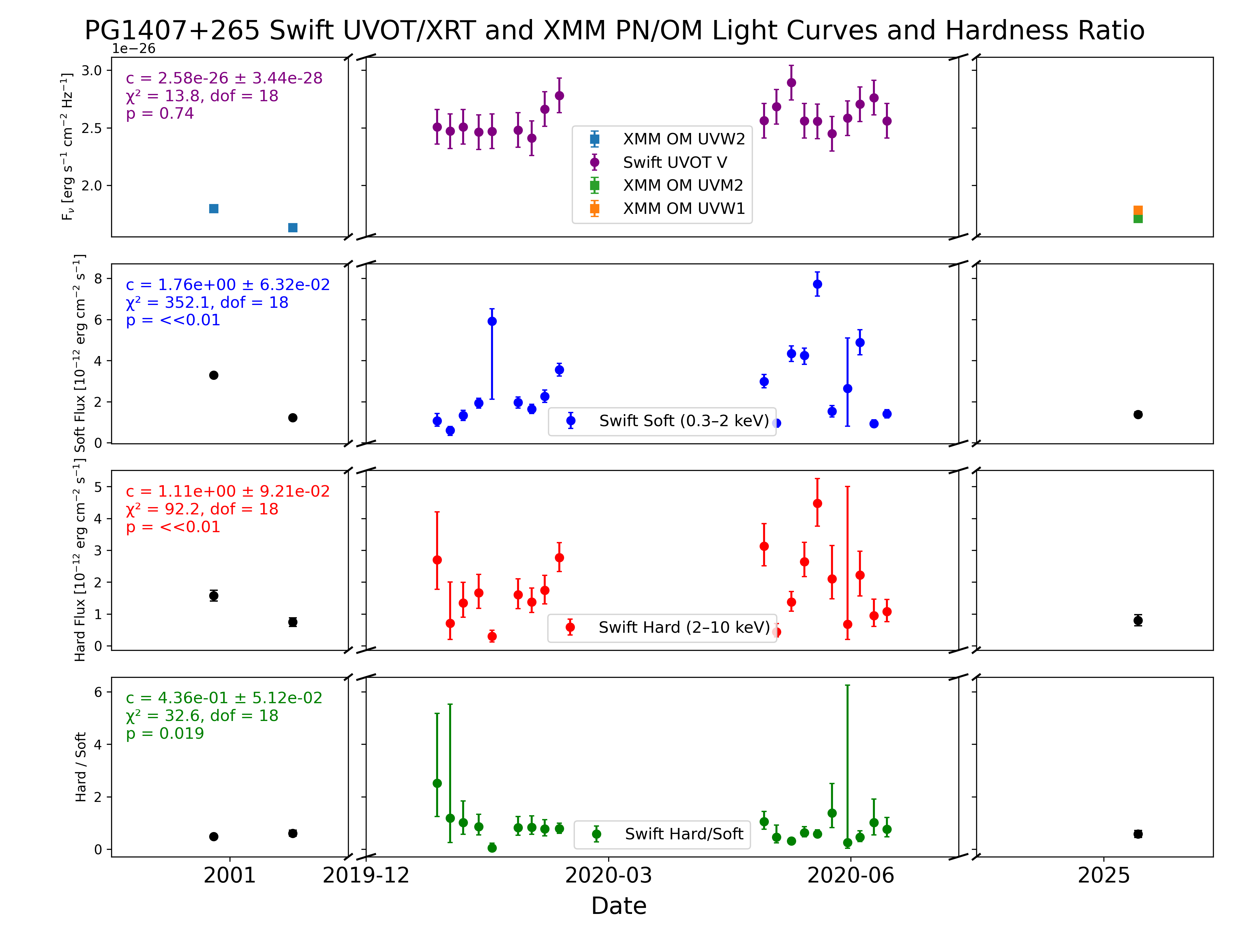}
    
        \label{fig:pg1407_uvot_xrt_lc}
    \end{subfigure}
    \hfill
    \begin{subfigure}[t]{0.9\textwidth}
        \centering
        \includegraphics[width=\textwidth]{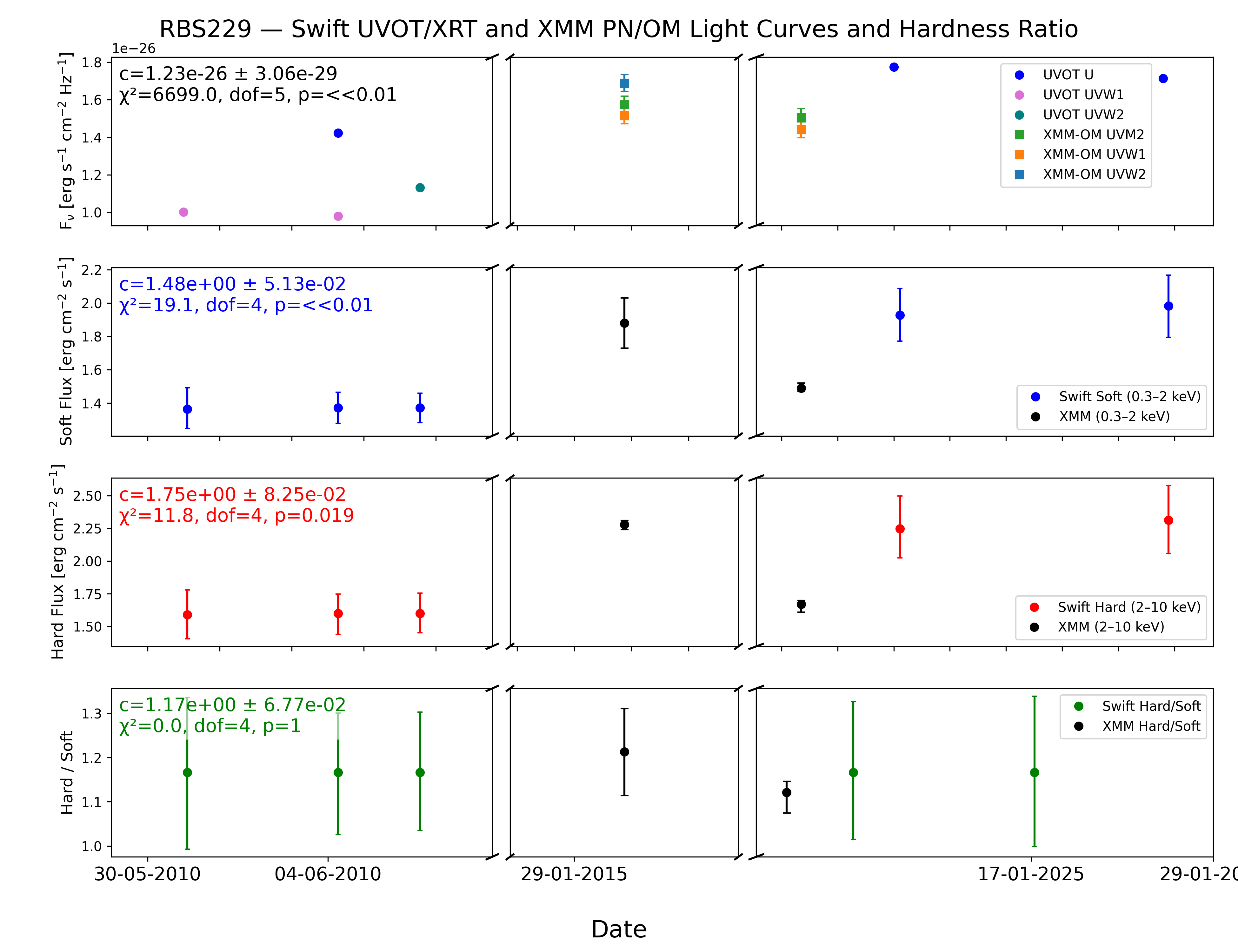}
       
        \label{fig:rbs229_uvot_xrt_lc}
    \end{subfigure}

    \caption{
    \textit{Swift}/UVOT, XRT, PN and OM light curves and hardness ratio.
    From top to bottom: UV flux ($F_\nu$), soft X-ray flux (0.3--2~keV),
    hard X-ray flux (2--10~keV), and hardness ratio (hard/soft).
    }
    \label{fig:uvot_xrt_lc}
\end{figure*}

\begin{figure*}[htbp]
    \centering

    \begin{subfigure}{0.5\textwidth}
        \centering
        \includegraphics[width=\linewidth]{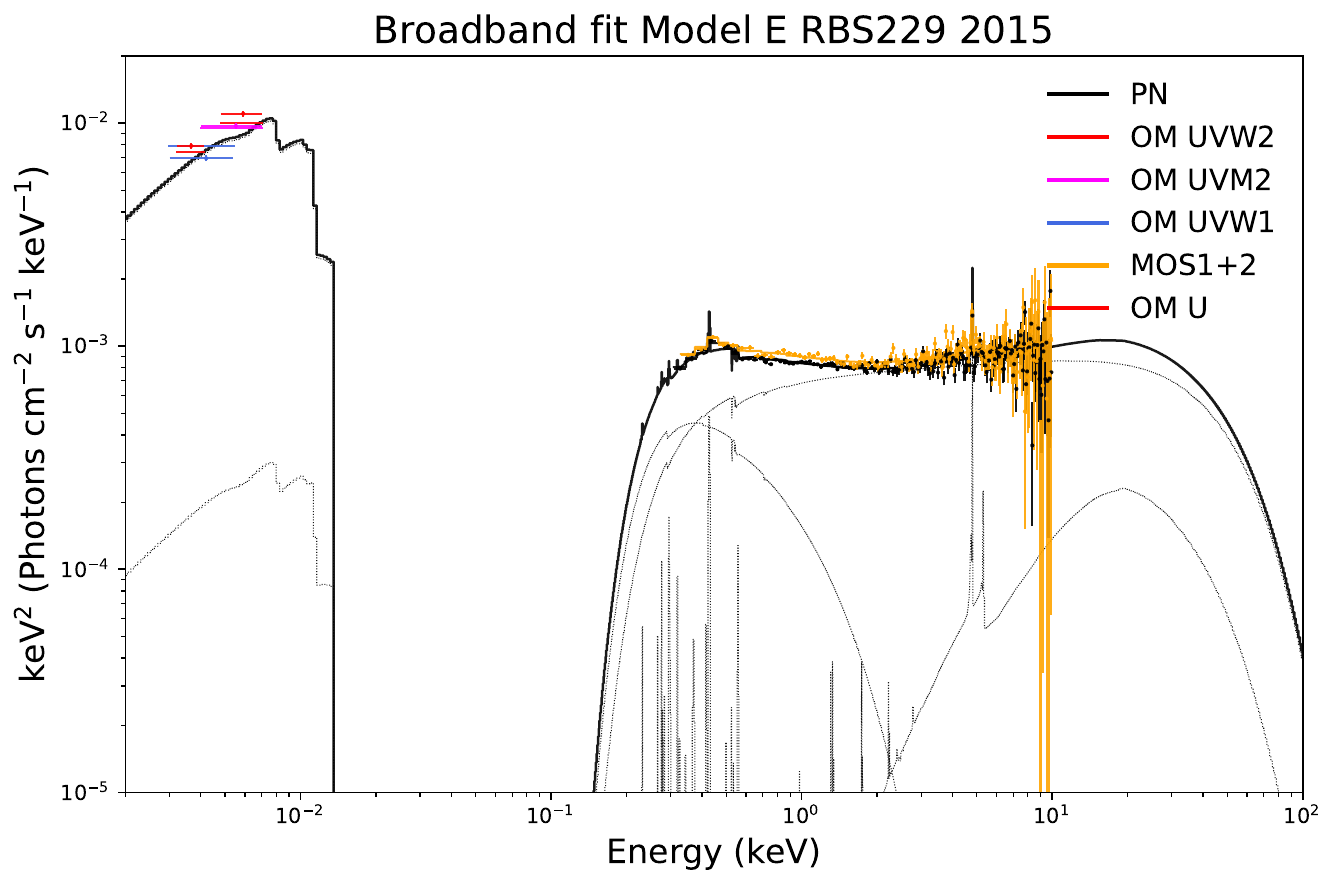}
    \end{subfigure}
    \hfill
    \begin{subfigure}{0.49\textwidth}
        \centering
        \includegraphics[width=\linewidth]{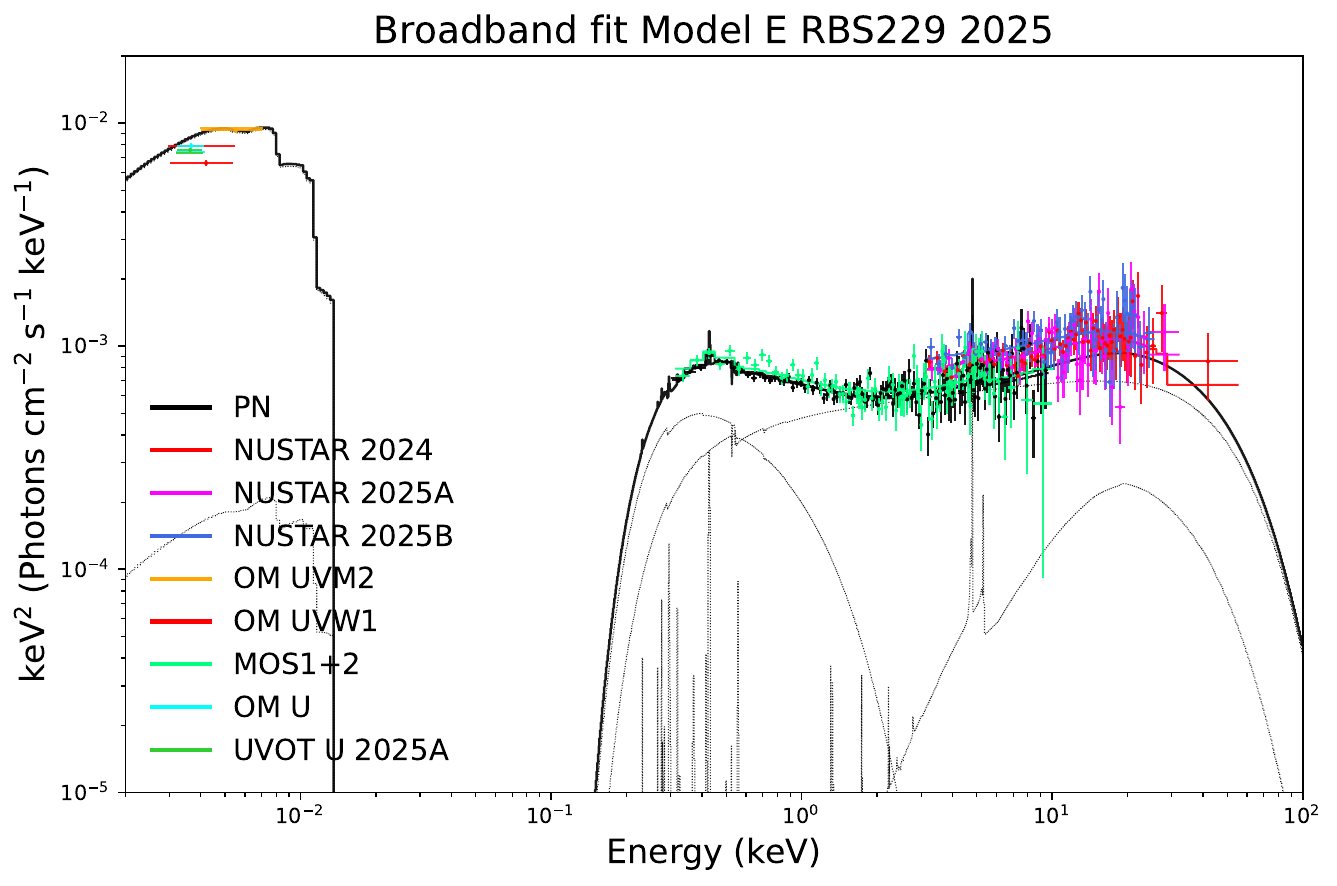}
    \end{subfigure}

    \vspace{0.5cm}

    \begin{subfigure}{0.49\textwidth}
        \centering
        \includegraphics[width=\linewidth]{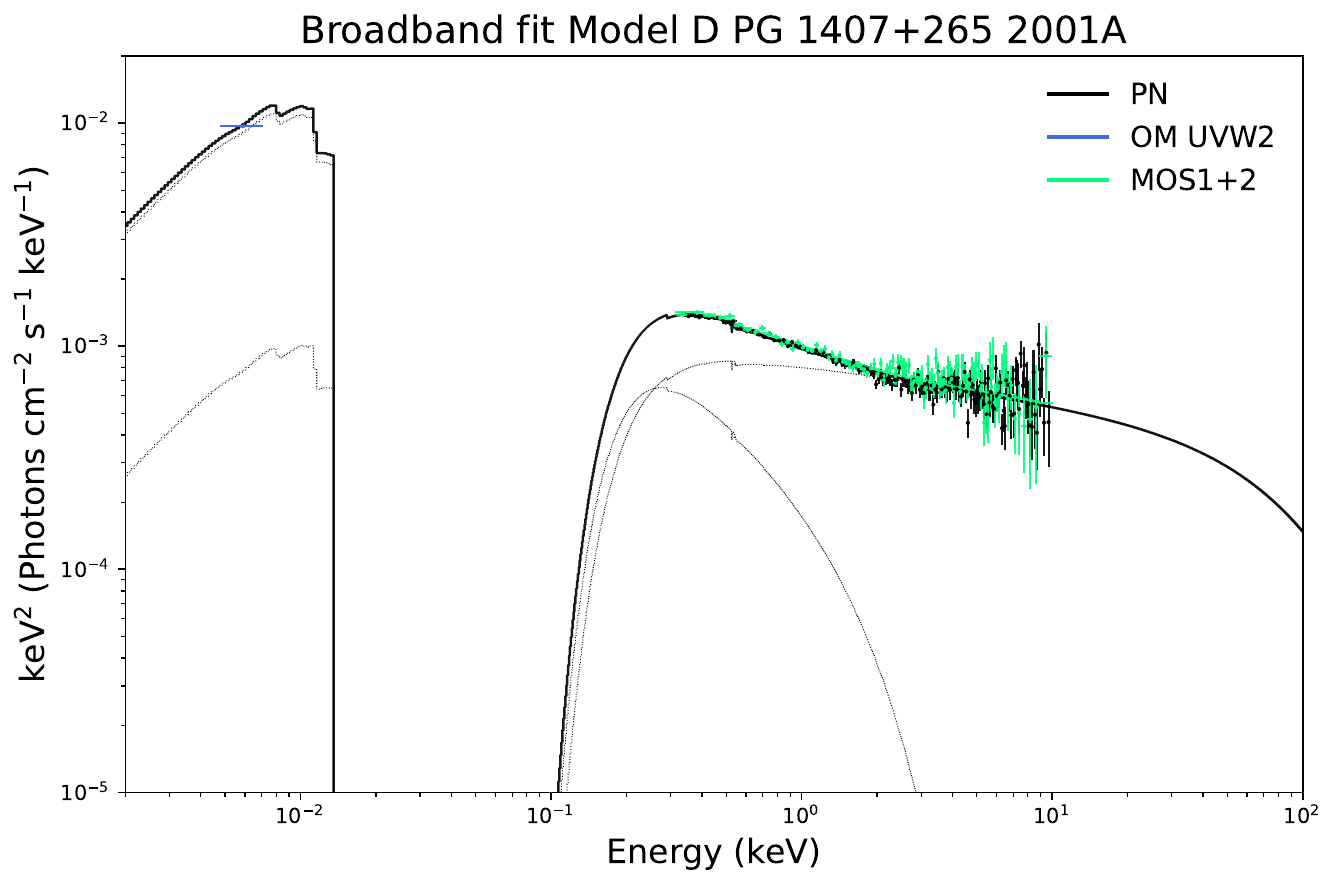}
    \end{subfigure}
    \hfill
    \begin{subfigure}{0.49\textwidth}
        \centering
        \includegraphics[width=\linewidth]{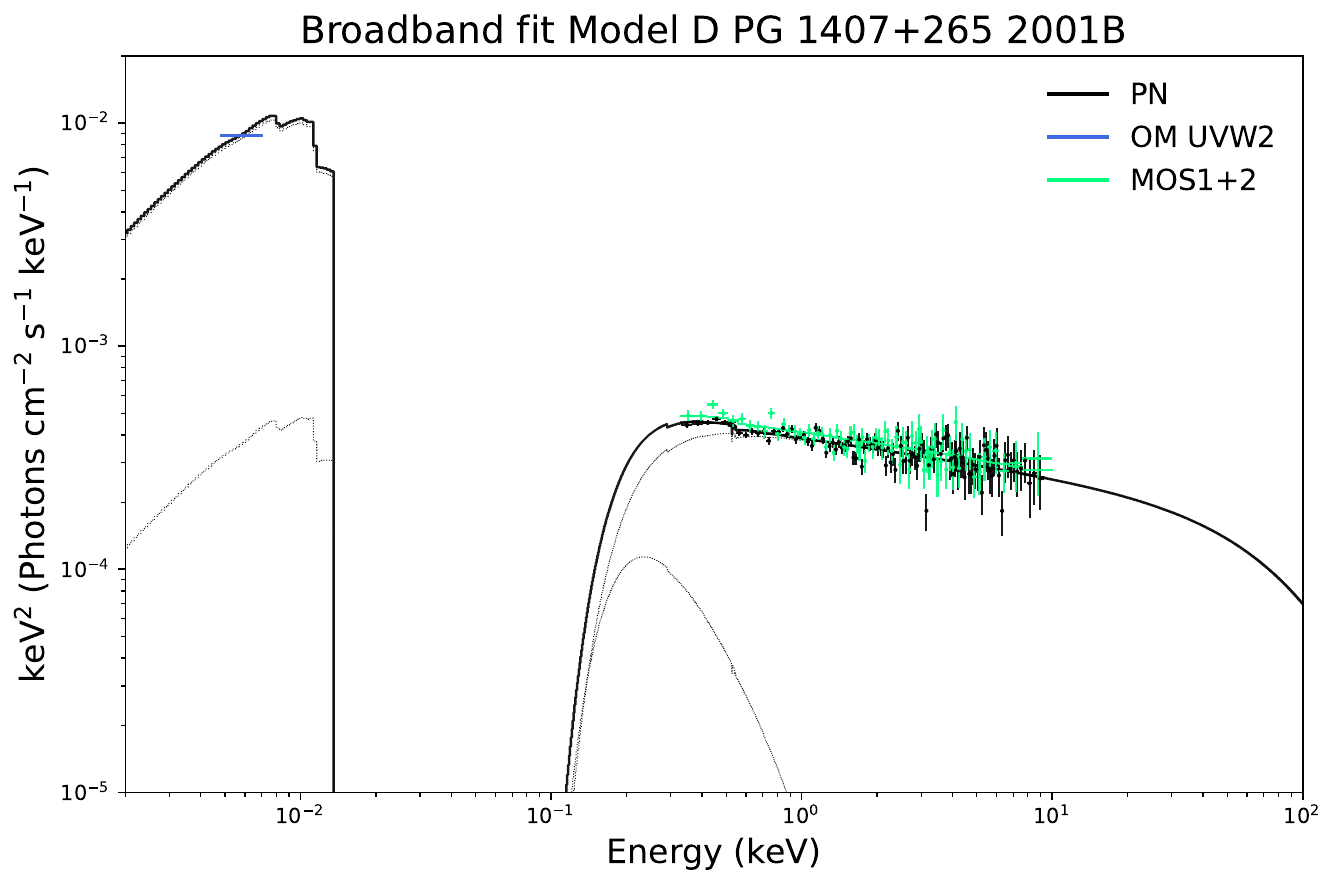}
    \end{subfigure}

    \vspace{0.5cm}

    \begin{subfigure}{0.49\textwidth}
        \centering
        \includegraphics[width=\linewidth]{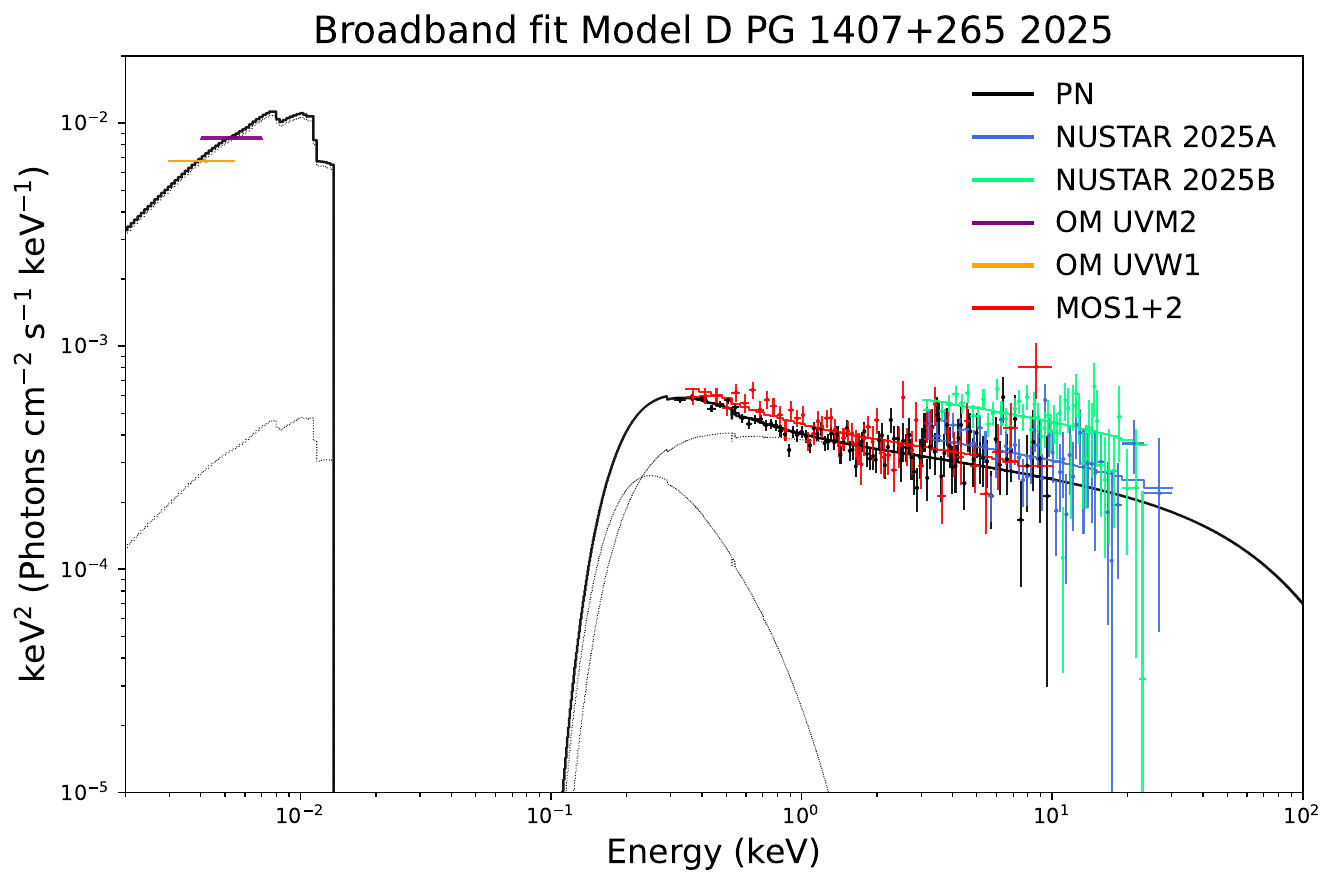}
    \end{subfigure}

   \caption{
Broadband X-ray/UV spectra fitted with the best-fit spectral model for each observation epoch.
The panels are ordered chronologically in time, showing the spectral evolution of RBS~229 (2015, 2025) and PG~1407+265 (2001A, 2001B, 2025).
For each epoch, the spectra from the different instruments are shown together with the corresponding best-fit model components.
}
    \label{fig:broadband_plots}
\end{figure*}

\end{appendix}

\end{document}